\documentclass[12pt]{article}
\usepackage[utf8]{inputenc}
\usepackage{amssymb}
\usepackage{pdfpages}
\usepackage{lscape}
\usepackage{natbib}
\usepackage{amsmath}
\usepackage{longtable}

\DeclareMathOperator{\bu}{\mathbf{u}}
\DeclareMathOperator{\bv}{\mathbf{v}}
\DeclareMathOperator{\bX}{\mathbf{X}}

\DeclareMathOperator{\ut}{\Tilde{u}}
\DeclareMathOperator{\bM}{\textit{M}}
\DeclareMathOperator{\bC}{\textit{C}}
\DeclareMathOperator{\bR}{\textit{R}}
\DeclareMathOperator{\bU}{\mathbf{U}}
\DeclareMathOperator{\bV}{\mathbf{V}}
\DeclareMathOperator{\bS}{\mathbf{S}}
\def\smt{{\mbox{\tiny T}}}

\usepackage[lined,boxed,linesnumbered]{algorithm2e}
	\DontPrintSemicolon
\usepackage{graphicx}
\usepackage{color}
\usepackage{url}

\usepackage{verbatim}
\usepackage[all]{xy}
\xyoption{arc}
\usepackage{xkeyval} 
\usepackage{tikz}
\usepackage{etoolbox}
\usepackage{multirow}
\usepackage{caption}
\usetikzlibrary{intersections}

\date{}
\begin{document}
 \title{Robust Integrative Biclustering for Multi-view Data }
\begin{tiny}
  \author{W. Zhang$^{\text{\sf 1}}$, C. Wendt$^{\text{\sf 2}}$, R. Bowler$^{\text{\sf 3}}$, C.  P. Hersh$^{\text{\sf 4}}$, and S. E. Safo$^{\text{\sf 1}}$\thanks{corresponding author email: ssafo@umn.edu}\\[4pt]
$^{\text{\sf 1}}$Division of Biostatistics, $^{\text{\sf 2}}$Division of Pulmonary, Allergy and Critical Care \\
 University of Minnesota, Minneapolis, 55455, USA \\
$^{\text{\sf 3}}$Division of Pulmonary, Critical Care and Sleep Medicine, \\ Department of Medicine, National Jewish Health, Denver, CO, USA  \\
$^{\text{\sf 4}}$Channing Division of Network Medicine, Brigham and Women's Hospital\\ Harvard Medical School, Boston, MA, USA}
\end{tiny}
\markboth%
{Zhang et.al.}
{Robust Integrative Biclustering}
    
  \maketitle
\def\spacingset#1{\renewcommand{\baselinestretch}%
{#1}\small\normalsize} 
\begin{abstract}
In many biomedical research, multiple views of data (e.g., genomics, proteomics) are available, and a particular interest might be the detection of sample subgroups characterized by specific groups of variables. Biclustering methods are well-suited for this problem as they assume that specific groups of variables might be relevant only to specific groups of samples. Many biclustering methods exist for identifying row-column clusters in a view but few methods exist for data from multiple views. The few existing algorithms are heavily dependent on regularization parameters for getting row-column clusters, and they impose unnecessary burden on users thus limiting their use in practice. We extend an existing biclustering method based on sparse singular value decomposition for single-view data to data from multiple views. Our method, integrative sparse singular value decomposition (iSSVD), incorporates stability selection to control Type I error rates, estimates the probability of samples and variables to belong to a bicluster, finds stable biclusters, and results in interpretable row-column associations. Simulations and real data analyses show that iSSVD outperforms several other single- and multi-view biclustering methods and is able to detect meaningful biclusters.  iSSVD is a user-friendly, computationally efficient algorithm that will be useful in many disease subtyping applications. \\

\end{abstract}

\maketitle

\spacingset{1}
\section{Introduction}
Biclustering (or two-way clustering, co-clustering, two-mode clustering) is a popular statistical method for simultaneously identifying groups of samples (rows) and groups of variables (columns) characterizing different sample groups. These clusters of rows and columns are known as biclusters. Biclustering methods are especially appealing for complex disease subtyping as they seek to identify homogeneous subgroups of people characterized by highly specific groups of biological features. A main limitation of one-way clustering algorithms  such as hierarchical or k-means clustering when applied to high-dimensional molecular data for disease subtyping is that cluster assignment of samples is based on the assumption  that all molecular features are relevant to the sample groups or disease subtypes. But specific groups of genes, for instance, may be co-regulated within one disease subtype, and not another subtype. In such cases, biclustering methods are well-suited. 

Generally speaking, a biclustering algorithm finds the associations between observations (rows) and attributes (columns) in a data matrix. More recently, because of the availability of three-dimensional data such as \textit{gene-sample-time} data in biomedical research, a number of tri-clustering methods have been proposed to identify homogeneous three-dimensional (3D) subspaces in a given 3D data set. As noted by Henriques and Madeira \cite{tri-cluster-review}, tri-clustering algorithms  face a number of major challenges such as robustness and efficiency. In this paper we focus on biclustering, which aims to identify row-column associations in a two-dimensional (2D) data matrix.

A number of biclustering methods have been proposed over the past two decades and they can be broadly categorized into four groups: a) combinatorial methods such as CTWC \citep{CTWC}, OPSM \citep{OPSM}, BIMAX \citep{BIMAX}, association analysis based RAP \citep{RAP}, COALESCE \citep{COALESCE}, QUBIC \citep{QUBIC}, and QUBIC2 \citep{qubic2}; b) probabilistic and generative methods such as SAMBA\citep{SAMBA}, FABIA \citep{FABIA}, BicMix \citep{bicmix}, COBRA \citep{convex-biclustering}, GBC \citep{GBC}, and plaid models \citep{plaid0}; c) matrix factorization approaches that include SSVD \citep{Lee}, S4VD \citep{Robust}, biclustering via non-negative matrix factorization \citep{NMF}, and BEM \citep{BEM}; and d) deep learning with neural networks such as AutoDecoder \citep{Autodecoder}. Other biclustering methods have been proposed to tackle specific problems such as missing data \citep{biclustering-missing}, heterogeneous and temporal medical data \citep{biclustering-hetreo-temporal-data},  and heterogeneous data from multiple views \citep{GBC}.

So far, most existing biclustering methods aim to detect row-column clusters in a single data matrix (i.e., single-view or data from one view). The integration of multi-view  data in biomedical research has garnered considerably interests nowadays, thanks to advancements in technology and data preprocessing \citep{multiview-review}. Such methods exploit the strengths in individual data as well as the overall dependency structure among multiple views. Biclustering methods that leverage the rich information in diverse but related data (e.g., genomics, proteomics) have potential to identify multidimensional view-specific features characterizing sample subgroups common to all views. Figure 1 is a pictorial illustration of biclustering for two views. For instance, by integrating genetic and clinical data, we leverage the strengths in molecular data and the advantages of clinical data to define homogeneous groups of people with common molecular information and  clinical factors likely contributing to disease outcome.  Yet, there are only few biclustering methods that have been developed for data from multiple views \citep{multibayes,multicoclustering,multisvd, GBC}. For data from two views, existing tri-clustering algorithms could be used to identify homogeneous subspaces. Bunte et al. \citep{multibayes}  developed a Bayesian approach for joint biclustering of data from multiple views that is based on group factor analysis. In Sun et al. (2014) \citep{multisvd}, a multi-view biclustering method based on sparse singular decomposition \citep{Lee} was proposed. The authors in Sun et al. (2015) \citep{multicoclustering} proposed a multi-view biclustering method that is based on low-rank matrix approximation and a proximal alternating linearized minimization algorithm was developed to solve their optimization problem. Recently, a Bayesian biclustering method for integrating data from multiple views that allow for each data to have a different distribution has been proposed\citep{GBC}.

\begin{figure}
    \centering
    \includegraphics[scale=0.8]{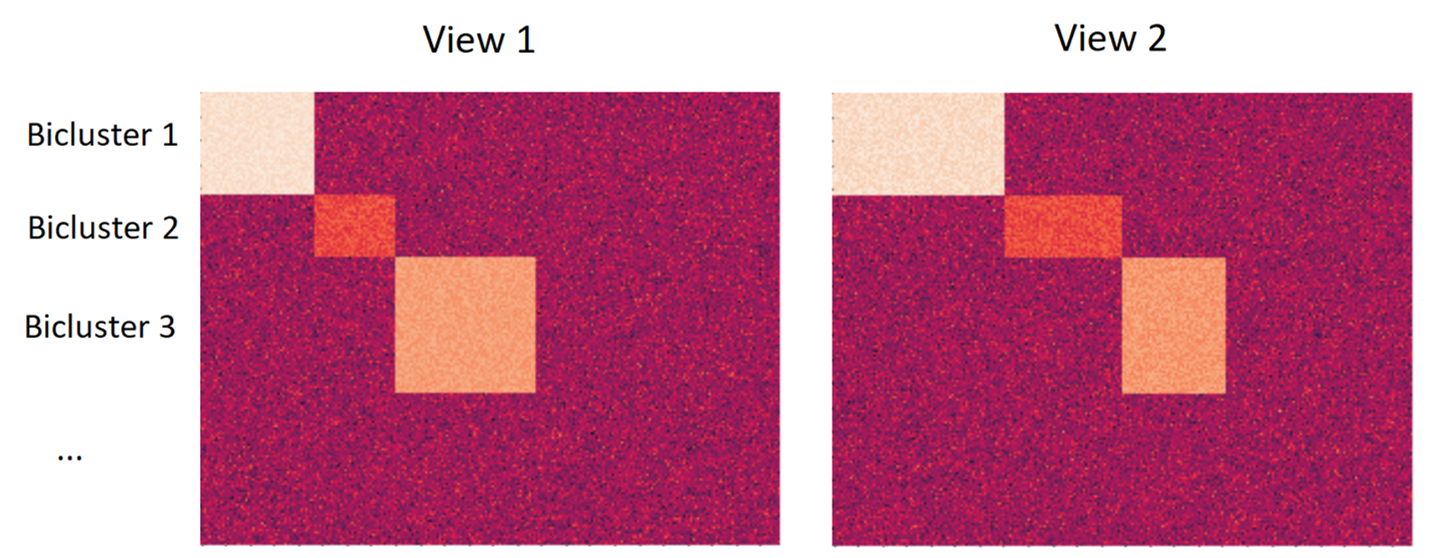}
    \caption{Pictorial illustration of integrative biclustering with two views. A bicluster is comprised of rows (or samples) and columns (or variables) from each view, with the samples in view 1 being the same as the samples in view 2. The samples in each bicluster can overlap or not. Similarly, the variables in each bicluster within a view can overlap or not. In this figure, the samples and variables in each bicluster do not overlap.}
    \label{fig:bicluster}
\end{figure}

The sparse singular value decomposition (SSVD) biclustering method for single-view data \citep{Lee}  obtains a sparse rank-one approximation of the data. To obtain the first bicluster, the authors  minimize the Frobenius norm between the data and the rank one approximation of the data while  regularizing both the left and right singular vectors using adaptive lasso penalties \citep{zou2006adaptive}. The degree of sparsity of the singular vectors depend strongly on the choice of the regularization parameters. Several techniques including Bayesian information criterion (BIC), Akaike Information criterion (AIC) and cross-validation have been proposed in the literature to select regularization parameters.  The Bayesian information criterion  was used to choose the optimal tuning parameters in Lee et al. (2010) \citep{Lee}. However, the authors in Sill et al. (2011) \citep{Robust} observed that the BIC oftentimes resulted in low degree of sparsity and they proposed to use stability selection \citep{meinshausen2010stability} to choose the regularization parameters and to determine stable biclusters. Stability selection is a subsampling procedure that was originally proposed  to select stable variables for penalized regression models and since  then it has been used successfully in other settings. This approach allows for choosing  penalization parameters and further controlling Type I error rates. 

In this article, we extend the SSVD method \citep{Lee} and the SSVD method with stability selection \citep{Robust} to identify biclusters in multi-view data. Our main contribution, compared to other multi-view biclustering methods \citep{multicoclustering,multisvd, GBC} is that we use stability selection to identify stable and robust subject and variable clusters.  We note two other contributions to existing biclustering methods that we believe are important. Compared to the methods in Sill et al. (2011) \citep{Robust}, Sun et al. (2014) \citep{multisvd}, and Sun et al. (2015) \citep{multicoclustering}, we estimate subsequent biclusters using the whole data. To ensure that each sample belongs to only one subject cluster, the aforementioned methods use only unclustered samples when deriving subsequent sample and column clusters. This is concerning since smaller sample sizes are used to estimate subsequent biclusters. We track samples clustered and assign weights that ensure that clustered samples have zero coefficients in subsequent biclusters. Also, we develop  efficient and user-friendly algorithms for the proposed method. We use extensive simulations to compare the performance of our proposed method in both biclusters detection and computational time. We apply the method to RNA sequencing and proteomics data from the Genetic Epidemiology of Chronic Obstructive Pulmonary Disease Study (COPDGene) \citep{regan2011genetic} to  identify subject and molecular clusters. The sample groups identified are compared across several demographic, clinical, and lung function variables. 

The rest of the paper is organized as follows.  In Section 2, we introduce  the proposed method.  In Section  3, we conduct simulation studies to assess the performance of our method in comparison with other methods in the literature. In Section 4, we apply our proposed method to a real data. We end with some brief discussion in Section 5.

\section{Methods}
Our primary goal is to define a biclustering method that leverages the wealth of information from diverse but related data to identify stable and robust sample clusters characterized by sample-specific variables. In this section, we first briefly summarize previous biclustering methods for one view that are based on singular value decomposition (SVD) which motivates our proposed work. Then, we extend previous work to multiple views. Finally, we incorporate the concept of stability selection to identify robust biclusters and control Type 1 error rates of falsely selecting samples and variables in a bicluster.

\subsection{Biclustering for a single view}\label{sec:biclusterone}
Let $\bX$ be a $n \times p$ data matrix where $n$ represents samples and $p$  variables. We assume that the data can be approximated by sparse rank $k$ ($k=1,\ldots,K$) left and right singular vectors $\bu_k$ and $\bv_k$ and corresponding singular value $s_k$, i.e., $\bX \approx \sum_{k=1}^{K}s_k\bu_k\bv_k^{\smt}$. Then, the non-zero entries in $\bu_k$ represent a  sample subgroup in the $k$th left singular vector. Similarly, the non-zero entries in $\bv_k$ represent a variable subgroup in the $k$th right singular vector. Together, these represent the $k$th bicluster. Lee et al. \citep{Lee} proposed to obtain the best sparse rank one approximation of the data by solving the following optimization problem:
 \begin{eqnarray} \label{ssvd} 
(\widehat{s}_1,\widehat{\bu}_1,\widehat{\bv}_1)  =   \min_{s_1,\bu_1,\bv_1}  \|\bX - s_1\mathbf{u_1}\mathbf{v_1}^T\|_F^2 + \lambda_{u_1}
\mathcal{P}_u(s_1\mathbf{u_1}) + \lambda_{v_1} \mathcal{P}_v(s_1\mathbf{v_1})\nonumber\\
\mbox{subject~to}\|\bu_1\|_2=1, ~~ \|\bv_1\|_2=1.
\end{eqnarray}
Here, the first term measures the approximation error using Frobenius norm of the difference;  $\mathcal{P}_i(\cdot), i=u,v$ are sparsity-inducing terms, and $\lambda_{u_1}$ and $\lambda_{v_1}$ control the level of sparsity. For fixed $\bu_1$, the authors minimize \eqref{ssvd} with respect to $\widetilde{\bv}_1 = s_1\bv_1$. Similarly,  for fixed $\bv_1$ they minimize \eqref{ssvd} with respect to $\widetilde{\bu}_1 = s_1\bu_1$. Adaptive lasso penalties \citep{zou2006adaptive} were used to induce sparsity on the vectors, i.e., 
    \begin{equation}
        \mathcal{P}_u(s_1\bu_1) = s_1\sum_{i=1}^{n} w_{1,i} |u_{1i}| \:\: \text{and} \:\: \mathcal{P}_v(s_1\bv_1) = s_1\sum_{j=1}^p w_{2,j} |v_{1j}|,
    \end{equation}
    where $w_{1,i}$'s and $w_{2,j}$'s are adaptive weights (obtained from the data itself). The adaptive lasso penalties become lasso penalties when $w_{1,i} = w_{2,j} = 1$, $\forall i,j$.
To obtain $\bu_1$ and $\bv_1$, Lee et al. \citep{Lee} iteratively solved the optimization problem \eqref{ssvd} until convergence. The estimated  singular value is then $\widehat{s}_{1}= \widehat{\bu}_1\bX\widehat{\bv}_1$. The reconstructed matrix, after convergence, given by $\widehat{s}_1\widehat{\bu}_1\widehat{\bv}_1^{\smt}$ is one sparse SVD layer.  The non-zero entries in $\widehat{\bu}_1$ and $\widehat{\bv}_1$ form the first bicluster. Subsequent biclusters may be obtained by a sparse rank one approximation of the deflated data ($\bX - \widehat{s}_1\widehat{\bu}_1\widehat{\bv}_1^{\smt}$), i.e., by repeatedly solving problem (\ref{ssvd}) using deflated data. 

\subsection{Biclustering for multiple views}\label{sec:biclustermulti}
We extend the biclustering problem to data from multiple views. Suppose that data are available from $D$ different views, and each view is arranged in an $n \times p^{(d)}$ matrix $\bX^{(d)}$, where the superscript $d$ correspond to the $d$th view. For instance, for the same set of $n$ individuals, matrix $\bX^{(1)}$ consists of RNA sequencing data and $\bX^{(2)}$ consists of proteomics data, for $D=2$ views. We wish to cluster the $p^{(d)}$ variables, and the $n$ subjects so that each subject subgroup is associated with specific variable subgroup of relevant variables. Similar to the method proposed by Liu et al. \citep{irPCA},  we posit an unobserved common factors $\bu_k$ and view-specific factors $\bv_k^{(d)}$ that approximate each view, i.e., $\bX^{(d)} \approx \sum_{k=1}^K s^{(d)}_k\bu_k \bv^{(d)\smt}_k$.  Here, $\bU=[\bu_1,\ldots,\bu_k]$ is an $n \times K$ matrix of common latent components that connects the $D$ views and induces dependencies across the views. Following problem (\ref{ssvd}), the best sparse rank one approximation for the $D$ views may be obtained by solving the optimization problem: 
   \begin{eqnarray} \label{iSSVD}
   \underset{\bu_1,\bv_1^{(d)}, d\in\{1,2,...,D\}}{\min}
       \sum_{d=1}^D\|\bX^{(d)} - s_1^{(d)}\bu_1\bv_1^{(d)\smt}\|_F^2+ \lambda_{u_1}\mathcal{P}_u(s_1\bu_1)+ \sum_{d=1}^D \lambda_{v_1}^{(d)}\mathcal{P}_v^{(d)}(s^{(d)}_1\bv^{(d)}_1)\nonumber\\
       \mbox{subject~to~}\|\bu_1\|_2=1, ~~ \|\bv_1^{(d)}\|_2=1.
   \end{eqnarray}
   In order to obtain sparse approximations, we use lasso penalties \citep{lasso} on both $\bu_1$ and $\bv^{(d)}_1$ for $d\in\{1,2,...,D\}$. That is,
   \begin{eqnarray}
       \mathcal{P}_u(s_1\bu_1) = \lambda_{u_1} \sum_{i=1}^n s_1|u_{1i}|
        \:\: \text{and} \:\: \mathcal{P}_v^{(d)}(s^{(d)}_1\bv_1^{(d)}) =  \sum_{j=1}^{p^{(d)}} \lambda_{v_1}^{(d)}s^{(d)}_1|v^{(d)}_{1j}|,
   \end{eqnarray}
   where $\lambda_{u_1}$ and $\lambda_{v_1}^{(d)}$ are regularization parameters. The penalties are sums of absolute values of the elements in the first singular vectors. For fixed $\bu_1$, minimizing \eqref{iSSVD} is equivalent to minimizing $D$ equations with similar forms,  with respect to $\widetilde{\bv}_1^{(d)} = s^{(d)}_1\bv^{(d)}_1$:
   \begin{eqnarray}
       \label{solvev}
 \widehat{\widetilde{\bv}}_1^{(d)} &=&\min_{\widetilde{\bv}_1^{(d)}}\|\bX^{(d)} - \bu_1\widetilde{\bv}_1^{(d)\smt}\|_F^2 + \lambda_{v_1}^{(d)} \sum_{j=1}^{p^{(d)}} |\widetilde{v}_{1j}^{(d)}| \nonumber \\
       &=&\min_{\widetilde{\bv}_1^{(d)}} \|\bX^{(d)}\|_F^2 + \sum_{j=1}^{p^{(d)}}((\widetilde{v}_{1j}^{(d)})^2 -2 \widetilde{v}_{1j}^{(d)} (\bX^{(d)^{\smt}}\bu_1)_j  +\lambda_{v_1}^{(d)} |\widetilde{v}_{1j}^{(d)}|).
   \end{eqnarray}
   The solution of this optimization problem may be obtained by soft-thresholding \citep{Lee}:
   \begin{equation}
       \widehat{\widetilde{v}}_{1j}^{(d)} = \text{sign}\{(\bX^{(d)^{\smt}} \bu_1)_j\} (|(\bX^{(d)^{\smt}} \bu_1)_j| - \lambda_{v_1}^{(d)}/2)_{+}.
   \end{equation} 
Then,  $\widehat{\widetilde{\bv}}_1^{(d)} =( \widehat{\widetilde{v}}_{11}^{(d)}, \ldots, \widehat{\widetilde{v}}_{1p}^{(d)})$ is an estimate for the product of the first right singular vector and the first singular value. As in Lee et al. (2010)\cite{Lee}, we obtain an estimated sparse rank one right singular vector for the $d$th view as ${\widehat{\bv}}_1^{(d)} = \widehat{\widetilde{\bv}}_1^{(d)}/\|\widehat{\widetilde{\bv}}_1^{(d)} \|_2$. The rank one sparse estimate for the left singular vector $\bu_1$ can be obtained in a similar way by concatenating the $D$ views. Let $\mathbf{X} = [\mathbf{X}^{(1)},\mathbf{X}^{(2)},...,\mathbf{X}^{(D)}] \in \Re^{n\times(p^{(1)}+p^{(2)}+...+p^{(D)})}$ be the concatenated data. Also,  let $\bv_1$ be a collection of the rank one right singular vectors for all $D$ views, i.e., $\bv_1 = (\bv_1^{(1)\smt},\ldots, \bv_1^{(D)\smt})^{\smt} \in \Re^{(p^{(1)} + \ldots+p^{(D)})}$. For $\bv_1$ fixed, we solve optimization problem \eqref{iSSVD} for $\bu_1$: 
   \begin{eqnarray}
       \label{solveu}
 \widehat{\widetilde{\bu}}_1 &=&      \min_{\widetilde{\bu}_1} \|\bX - \widetilde{\bu}_1\bv^{\smt}_1\|_F^2 +\lambda_{u_1} \sum_{i=1}^n |\ut_i| \nonumber \\
 &=&\min_{\widetilde{\bu}_1}\|\bX\|_F^2 + \sum_{i=1}^n((\widetilde{u}_{1i})^2 -2 \widetilde{u}_{1i} (\bX\bv_1)_i  +\lambda_{u_{1}} |\widetilde{u}_{1i}|)
   \end{eqnarray}
where $\widetilde{\bu}_1 = s_1\bu_1$ and $\bX$ is the concatenated data set. Using soft-thresholding again, the component-wise solution of \eqref{solveu} is 
\begin{equation}
     \widehat{\widetilde{u}}_{1i} = \text{sign}\{(\bX\bv_1)_i\} (|(\bX \bv_1)_i| - \lambda_{u_1}/2)_{+}.
\end{equation}
As before, the corresponding sparse rank one left singular vector is ${\widehat{\bu}}_1 =\widehat{\widetilde{\bu}}_{1}/ \|\widehat{\widetilde{\bu}}_{1}\|_2$. Subsequent update of $s_1^{(d)}$ is given by $\widehat{s}_1^{(d)}= \widehat{\bu}_1^{\smt}\bX^{(d)}\widehat{\bv}_1^{(d)}$. Then we deflate the $d$th data as $\bX^{(d)} - \widehat{s}_1^{(d)}\widehat{\bu}_1\widehat{\bv}_1^{(d)T}$ for $d \in \{1,2,...,D\}$.
For subsequent biclusters, we repeatedly solve problems (\ref{solvev}) and (\ref{solveu}) using deflated data, i.e. we find sparse rank one approximations of the deflated data. 

The regularization parameters, $\lambda_{u_1}$ and $\lambda_{v_1}^{(d)}$ control the degree of sparsity (i.e., the number of non-zero elements in $\bu_1$ and $\bv_1^{(d)}$ must be chosen). Lee et al. \citep{Lee} proposed a Bayesian Information Criterion \citep{BIC} to obtain the optimal regularization parameters. Sill et al. \citep{Robust} proposed to use stability selection techniques for  robust biclusters. Similar to that, we choose the regularization parameters using stability selection \citep{meinshausen2010stability}. 

\subsection{Multi-view biclustering with stability selection}\label{sec:biclustermultiss}
Stability selection method, proposed by Meinshausen and Bühlmann\citep{meinshausen2010stability}, has been used for variable selection \cite{stabilityregression} problems such as regularized regression and even for sparse singular value decomposition \citep{Robust}. Stability selection essentially combines resampling  with variable selection so that the probability that a variable is selected is based on its relative frequency. Meinshausen and Bühlmann\citep{meinshausen2010stability} provide a theoretical justification to show that by selecting variables based on the maximum of these probabilities, the  Type 1 error rates of falsely selecting variables is controlled. For completeness sake, we briefly summarize the stability selection method and we describe how we use it in our application.  

We consider estimating the left singular vector $\bu_1$ and inferring the non-zero coefficients or identifying samples that form a sample cluster. We subsample variables in each view  $I$ times without replacement, while ensuring that each view contains the same set of samples. For each regularization parameter $\lambda_{u_1}$ from a set of regularization parameters $\Lambda_{u_1}$, we solve the optimization problem (\ref{solveu}) for each subsampled data set. Each $\lambda_{u_1} \in \Lambda_{u_1}$ leads to a different set of non-zero coefficients. We estimate the selection probability for  each sample $i=1,\ldots,n$ as the number of times sample $i$ is selected from $I$ applications of equation (\ref{solveu}) for a fixed $\lambda_{u_1}$. Denote the selection probability corresponding to $\lambda_{u_1}$ for sample $i$ as $\widehat{\Pi}_i^{\lambda_{u_1}}$. Then for an arbitrary threshold, $\pi_{thr}$, the stable path for $\bu_{1}$, $\widehat{\mathcal{S}}^{\text{stable}}_{u_1}$, (the set of stable samples) is the set of non-zero coefficients with selection probabilities at least $\pi_{thr}$. Essentially, samples with high selection probability are kept, and those with low selection probabilities are disregarded. Then, given the union of the selected samples  from all $\lambda_{u_1} \in \Lambda_{u_1}$, we can estimate the average number of selected samples (i.e., non-zero coefficients) for the regularization region $\Lambda_{u_1}$, denoted as $q_{\Lambda_{u_1}}$. From Theorem 1 in Meinshausen and Bühlmann (2010) \citep{meinshausen2010stability}, the expected number of falsely selected samples, $E({\bu_1})$ with stability selection  is bounded by:
\begin{equation}
    E({\bu_1}) \le \frac{1}{2\pi_{thr} -1}\frac{q_{\Lambda_{u_1}}^2}{n}.
\end{equation}
Thus, by reducing the average number of selected samples (i.e.,  $q_{\Lambda_{u_1}}$) or by increasing the threshold $\pi_{thr}$, we reduce the expected number of falsely selected samples or the per-family error rate, or if we divide by $n$, the per-comparison error rate (PCER) \citep{typeI}.  It is noted by the authors that the  threshold value range  $\pi_{thr}=[0.6,0.9]$ tend to yield similar results. For  $\pi_{thr}$ fixed, if we choose the average number of selected samples $q_{\Lambda_{u_1}}$ to be at most $e_{\Lambda_{u_1}}=\sqrt{E({\bu_1})(2\pi_{thr} -1)n}$ (and hence the regularization region $\Lambda_{u_1}$), we control the family-wise error rate for some $E({\bu_1})$. Following ideas in Sill et al. (2011) \citep{Robust}, we estimate $\widehat{\widetilde{\bu}}_{1}$ with the smallest regularization parameter value in the regularization region that ensures that $q_{\Lambda_{u_1}} \le e_{\Lambda_{u_1}}$. Thus, the component-wise estimate for $\widehat{\widetilde{\bu}}_{1}$ is given by: 
\begin{equation}\label{eqn:compu}
           \widehat{\widetilde{u}}_{1i} = \text{sign}\{(\bX\bv_1)_i\} (|(\bX \bv_1)_i| - \lambda_{\min_{u_1}}/2)_{+}.
\end{equation}
\indent We estimate the right singular vectors $\bv_1^{(d)}$ and infer the non-zero coefficients or variables that form a variable cluster in a similar way. Specifically, for each possible $\lambda_{v_1}^{(d)}$, we draw $J$ subsamples without replacement and we estimate the selection probabilities for each variable $j = 1,\dots,p^{(d)}$ as the number of times variable $j$ is selected from $J$ applications of equation \ref{solvev}. Denote the selection probability corresponding to $\lambda^{(d)}_{v_1}$ for variable $j$ as $\widehat{\Pi}_j^{\lambda_{v_1}^{(d)}}$. Given a threshold $\pi_{thr}$ and the desired Type 1 error value $E({\bv_1^{(d)}})$, we obtain the regularization region $\Lambda_{v_1^{(d)}}$ such that $q_{\Lambda_{v_1^{(d)}}} \le e_{\Lambda_{v_1^{(d)}}}$, where $e_{\Lambda_{v_1^{(d)}}} = \sqrt{E({\bv_1^{(d)}})(2\pi_{thr} -1)p^{(d)}}$. Then the stable set for $\bv^{(d)}_1$, $\widehat{\mathcal{S}}^{\text{stable}}_{v_1^{(d)}}$, are the  non-zero coefficients or variables with selection probabilities at least $\pi_{thr}$. Given the smallest regularization parameter value in the regularization region, the component-wise estimate for $\widehat{\widetilde{\bv}}_{1}^{(d)}$ is given by: 
   \begin{equation}\label{eqn:compv}
       \widehat{\widetilde{v}}_{1j}^{(d)} = \text{sign}\{(\bX^{(d)^{\smt}} \bu_1)_j\} (|(\bX^{(d)^{\smt}} \bu_1)_j| - \lambda_{\min_{v_1}}^{(d)}/2)_{+}.  
   \end{equation}
At convergence, the components of $\widehat{\bv}_1^{(d)}$ become $\widehat{v}_{1j}^{(d)}=\mathbf{1}(j \in \widehat{\mathcal{S}}^{\text{stable}}_{v_1^{(d)}})\widehat{v}_{1j}^{(d)}$, where $\mathbf{1}(\cdot)$ is an indicator function. Similarly, the components of $\widehat{\bu}_1$ become $\widehat{u}_{1i}=\mathbf{1}(i \in \widehat{\mathcal{S}}^{\text{stable}}_{u_1})\widehat{u}_{1i}$.

The algorithm iterates between $\bu_1$ and $\bv_1^{(d)}, d=1,\ldots,D$ until there is convergence. Refer to Algorithm \ref{alg::PGD} for more details.  For convergence, we estimate 1) the relative difference between the objectives ($\sum_{d=1}^D\|\bX^{(d)} - \widehat{s}_1^{(d)}\widehat{\bu}_1\widehat{\bv}_1^{(d)\smt}\|_F^2)$ at previous and current iterations, and 2) $\max(\|\bu_{1}-\widehat{\bu}_{1}\|^2, \min(\|\bv_{1}^{(1)}-\widehat{\bv}_{1}^{(1)}\|^2,\ldots,\|\bv_{1}^{(D)}-\widehat{\bv}_{1}^{(D)}\|^2))$. The algorithm converges if  either 1) or 2) is less than a pre-specified threshold  (e.g., 0.0001). 

We note that our method, referred to as integrative sparse singular value decomposition (iSSVD), is different from the one proposed in  \cite{irPCA}  because we are concerned with the problem of simultaneously identifying row (sample) and column (variable) clusters. As such, we regularize both $\bu_1$ and $\bv^{(d)}_1$, while Liu et al. \citep{irPCA} regularize $\bv^{(d)}_1$ and apply $k$-means clustering on $\bu$ after convergence. Our approach allows us to define  subgroups in rows and columns of our data simultaneously and makes it appealing to identify sample subgroups characterized by specific groups of  variables. Further, compared with several existing methods \citep{irPCA, icluster,multisvd,multicoclustering}, we use stability selection to identify stable and robust sample and variable clusters, while controlling for Type 1 error of falsely selecting samples and variables in a bicluster. In addition, to ensure that each sample belong to only one subject cluster, the authors in  \cite{multisvd,multicoclustering} and \cite{Robust} proposed to use only unclustered samples when estimating subsequent biclusters. This is concerning to us since smaller sample sizes are used to estimate subsequent biclusters. Instead, we track samples that are clustered and we assign weights to ensure that those samples have zero coefficients in subsequent biclusters. We do the same for variable clusters if it is desired to have non-overlapping variable clusters. Of note, concatenating the $D$ views and applying the biclustering method with stability selection proposed by Sill et al. \cite{Robust} assumes that the regularization parameters are the same for each view, i.e., $\lambda_{v_{1}}=\lambda_{v_{1}}^{(1)}=\cdots=\lambda_{v_{1}}^{(D)}$. This assumption may result in choosing tuning parameters that are either too small or too large  for a particular view;  this  can lead to a solution that is trivial or not sparse, and can inflate Type 1 error (refer to Supplementary Material for more details). 
\\
\noindent \textbf{Remark 1 Point-wise Control:} 
Searching for a plausible range of tuning parameters is computationally demanding. For computational efficiency, we implemented the point-wise control methods \citep{meinshausen2010stability, Robust}.  Specifically, we considered a path that shortens the time to find the optimal $\lambda_{u_{1}}$ or $\lambda_{v_{1}}^{(d)}$. We adopted the point-wise error control approach implemented in s4vd \citep{Robust} and expanded it to be feasible for multi-view data. For example, if solving for $\widehat{\bu}_1$, we can look for a regularization path with a single tuning parameter $\Lambda_{\bu_1} = \{\lambda_{u_1}\}$ and draw subsamples $I$ to calculate the average number of selected coefficient $q_{\Lambda_{\bu_1}}$, then we can  estimate the selection threshold by \begin{equation}
    \pi_{thr} = \dfrac{1}{2} \left( \dfrac{q^2_{\Lambda_{\bu_1}}}{E(\bu_1)n}+1 \right).
\end{equation} 
We define a region for the threshold to be $[\pi_{min}, \pi_{max}]$ and we search for a $\pi_{thr}^{'}$ that falls into this range. We start from the median value of the tuning parameter range, and this range is updated based on the reconstructed threshold. Next calculation uses the median value of the new tuning range and this continues until the reconstructed threshold satisfies the aforementioned range. Thus, instead of calculating the entire stability paths in each iteration, the algorithm finds appropriate parameters with fewer calculations. We incorporated this algorithm into iSSVD and compared the run time with s4vd.

\noindent \textbf{Remark 2 Choosing the number of biclusters:} To choose the  number of biclusters (i.e., $K$), we implemented the following approaches: a.) For each view, we calculate the proportions of variation explained by its singular values and select the number of singular values associated with the variation proportion that is larger than a threshold (for example larger than 70 percent); then we set the number of biclusters to be the maximum number of singular values plus one. b.) The user can specify the number of biclusters to be detected beforehand. Then the algorithm will set the maximum number of biclusters to be the smaller number from either a.) or b.). Table S2.3 gives the proportion of the true number of biclusters detected using the first criteria. 

\begin{scriptsize}
\begin{algorithm}[]
	\DontPrintSemicolon
	\SetKwComment{tcp}{$\triangleright\ $}{}
	\textbf{Input}: $\bX^{(d)}$, $K$ (optional), Type 1 errors $E(\bu)$ and $E(\bv^{(d)})$, $\pi_{thr}, ~~ d=1,\ldots,D$ \;
	
	\textbf{Output}: $\widehat{\mathbf{U}}=[\widehat{\bu}_1,\ldots,\widehat{\bu}_K]$, $\widehat{\mathbf{V}}^{(d)}=[\widehat{\bv}_1^{(d)},\ldots,\widehat{\bv}_K^{(d)}]$, $\kappa_u =\{k \in K \}_{i=1}^{n}$, $\omega_v^{(d)} =\{j \in K \}_{j=1}^{p^{(d)}}$   \;
\tcp*{{\small $\kappa_u$ and $\omega_v^{(d)}$ are vectors of bicluster membership. Samples that are not clustered will form a zero cluster.}}	
	\textbf{Initialize}: Apply standard SVD to the concatenated data $\bX$. Let $\{s_1,\bu_1,\bv_1\}$ be the first SVD triplet. (Note: $\bv_1 \in \Re^{p^{(1)}+\ldots+p^{(D)}}$). \;
     \Repeat{ convergence}{
     	\textbf{Solve for $\widehat{\widetilde{\bu}}_1$, and hence $\widehat{\bu}_1$} \;
	 \Begin{ 
	 (a) Draw subsamples $I$ and estimate $\widehat{\Pi}^{\lambda_{u_1}}_i$. Define $\Lambda_{u_1}$ such that $q_{\Lambda_{u_1}} \le e_{\Lambda_{u_1}}$. \;
	 (b) Estimate the set of non-zero sample coefficients, $\widehat{\mathcal{S}}^{\text{stable}}_{u_1}$ \;
	 (c) Solve for $\widehat{\widetilde{\bu}}_1$ using equation (\ref{eqn:compu}). Let $s_{u_{1}} =\|\widehat{\widetilde{\bu}}_1\|_2$, and $\widehat{\bu}_1 = \widehat{\widetilde{\bu}}_1/s_{u_{1}}$ \;
	 (d) Set  $\bu_1=\widehat{\bu}_{1}/\|\widehat{\bu}_{1}\|_2$
   	 }
   	 \For{$d=1,\ldots,D$}{ 	
   	 \textbf{Solve for $\widehat{\widetilde{\bv}}_1^{(d)}$, and hence $\widehat{\bv}_1^{(d)}$} \;
	 \Begin{ 
	 (a) Draw subsamples $J$ and estimate $\widehat{\Pi}^{\lambda_{v_1}^{(d)}}_j$. Define $\Lambda_{v_1^{(d)}}$ such that $q_{\Lambda_{v_1^{(d)}}} \le e_{\Lambda_{v_1^{(d)}}}$. \;
	 (b) Estimate the set of non-zero variable coefficients, $\widehat{\mathcal{S}}^{\text{stable}}_{v_1^{(d)}}$ \;
	 (c) Solve for $\widehat{\widetilde{\bv}}_1^{(d)}$ using equation (\ref{eqn:compv}). Let $s_{v_{1}}^{(d)} =\|\widehat{\widetilde{\bv}}_1^{(d)}\|_2$, $\widehat{\bv}_1^{(d)} = \widehat{\widetilde{\bv}}_1^{(d)}/s_{v_{1}}^{(d)}$ \;
	 (d) Set $\bv_1^{(d)}=\widehat{\bv}_{1}^{(d)}/\|\widehat{\bv}_{1}^{(d)}\|_2$; 
   	 }
   	 }
   		}
   	\textbf{Set} $\widehat{s}_1^{(d)}=\widehat{\bu}_1^{\smt}\bX^{(d)}\widehat{\bv}_1^{(d)}$;~~$\widehat{v}_{1i}^{(d)}=\mathbf{1}(i \in \widehat{\mathcal{S}}^{\text{stable}}_{v_1^{(d)}})\widehat{v}_{1i}^{(d)}$;  $\widehat{u}_{1i}=\mathbf{1}(i \in \widehat{\mathcal{S}}^{\text{stable}}_{u_1})\widehat{u}_{1i}$\;
   	\textbf{For subsequent biclusters} repeat steps 3 to 22 using deflated data. \;
   	\textbf{Stop} if either $\widehat{\mathcal{S}}^{\text{stable}}_{v_1^{(d)}} = \emptyset$ or $\widehat{\mathcal{S}}^{\text{stable}}_{u_1}=\emptyset$
\caption{Integrative Sparse Singular Value Decomposition Algorithm with Stability Selection (iSSVD) }
\label{alg::PGD}
\end{algorithm}
\end{scriptsize}

\newpage
\section{Simulations}
We consider two main scenarios to assess the proposed method in detecting biclusters from multi-view data. In both Scenarios, we simulate two views $\mathbf{X}^{(1)}$ and $\mathbf{X}^{(2)}$. In Scenario One, we allow for some samples to not belong to any sample cluster. In Scenario Two,  each sample belongs to a bicluster. This Scenario is especially relevant in disease subtyping where it is important that each sample belongs to only one sample cluster.  In each Scenario, we generate  fifty Monte Carlo datasets for each view. The parameter settings used can be found in supplementary material Table S1.5.
\subsection{Scenario One}
In this Scenario, data matrices $\bX^{(1)} \in \Re^{n \times p^{(1)}}$ and $\bX^{(2)}  \in \Re^{n \times p^{(2)}}$ are generated, respectively, as
\begin{equation}
    \bX^{(1)} = \mathbf{U}\bS\bV^{(1)\smt} + \mathbf{E}^{(1)}, ~~~    \bX^{(2)} = \mathbf{U}\bS\bV^{(2)\smt} + \mathbf{E}^{(2)}.
\end{equation}
Hence,  the concatenated data is $\bX = [\bX^{(1)}, \bX^{(2)}] \in \Re^{n \times (p^{(1)} +p^{(2)})}$. We set the number of biclusters $K=4$ and the dimensions of data to be $n=100$ and $p^{(1)}=p^{(2)} = 1,000$.   Each bicluster  has 10 rows and 100 columns. As such, there are only 40 samples that belong to the sample clusters with the remaining 60 samples not belonging to any cluster. Similarly, there are 400 signal variables characterizing the sample clusters, and the remaining 600 variables are noise. The left singular matrix $\bU \in \mathbb{R}^{100\times 100}$ is the common left singular matrix for the two views. 
Since we design four integrative biclusters, we randomly select 10 rows in each column of the first four columns in matrix $\bU$ and assign data values generated from a uniform distribution $U(0.5,1)$, while ensuring there is no overlapping samples. The remaining $n-40$ rows in the first four columns  are assigned zero values. For the entries of the remaining $n-K$ columns, we use data generated from the normal distribution with mean 0 and variance 1. We obtain the right singular matrix for each of the two views,  $\bV^{(d)} \in \mathbb{R}^{1000 \times 100}$, as follows. We randomly select 100 rows in each column of the first 4 columns in matrix $\bV^{(d)}$ and fill the corresponding elements from the uniform distribution $U(0.5,1)$, while also ensuring there are no overlapping rows (correspondingly no overlapping variables characterizing the sample clusters). 
As before, we fill out the entries of the remaining columns in $\bV^{(d)}$ with data generated from the  standard normal distribution. For the singular values $\bS$, we set the first four entries to 27, 20, 18 and 10  respectively, and the rest with small values $\epsilon$. Thus, $\bS \in \mathbb{R}^{100\times100} = diag(27,20,18,10,\epsilon,...,\epsilon)$ where $\epsilon = 0.3$; these singular values were chosen to make the biclusters detectable. Therefore, each view  is reconstructed as $\bX^{(d)} = \bU\bS\bV^{(d)\smt}, d=1,2$.  We then add random noise generated from a normal distribution with mean 0 and variance $\sigma^2$ to each view. We will assess the performance of the method for small to large variances. Since the two views  often tend to have different scales in real data analysis, we consider different scalings for $\bX^{(1)}$, $\bX^{(2)}$ as done in \cite{irPCA}.

\subsubsection{Case 1}
In this case, the two views  have different scales. Specifically, the concatenated data is of the form $\bX = [\bX^{(1)},s\bX^{(2)}]$ where $s \in \{1,2,5,10\}$; this allows us to investigate the performance of the proposed and existing methods in situations of unbalanced scales. This is commonly the case in multi-view data and it can be challenging for single-view methods.

\subsubsection{Case 2}
Case 2 is similar to Case 1 but we fix the scalar $s$ and study the performance of a method under different levels of noise. We vary the standard deviation $\sigma \in \{0.1,0.2,...,1\}$.

\subsubsection{Case 3}
In Case 3 we expand the dimensions of data to $n=500$ and $p^{(1)}=p^{(2)}=10,000$. Here, each bicluster has 50 rows and 200 columns. The remaining $n-200$ rows do not belong to any cluster, and the remaining $p^{(d)}-800$, $d=1,2$, variables are noise. We use this case to study the computational efficiency when $p^{(d)} \gg n$. \\

\subsection{Scenario Two}
In the first Scenario, we allowed for sample overlaps. In this scenario, each sample belongs to a bicluster. As before, we have two views  and they have dimensions $n=200$, $p^{(1)}=p^{(2)}=1,000$. There are four integrative biclusters for each view; each sample cluster has 50 samples, each variable cluster has 100 variables and the remaining $p^{(d)}-400$ variables are noise. The singular matrices are $\bU \in \Re^{n \times K}$, $\bS \in \Re^{K \times K}$ and $\bV^{(d)} \in \Re^{p^{(d)} \times K}$, $d=1,2$ and $K=4$. The fifty entries for each column in  $\bU$ are generated from the uniform distribution U(0.5,1), while ensuring that there are no overlapping entries. The remaining 150 entries in $\bU$ for each $k$ column is set to $0$. For the right singular matrix, $\bV^{(d)}$, $d=1,2$, we generate 100 entries for each column from U(0.5,1). The remaining $p^{(d)}-100$ entries in each column are  set to zero. For the singular values, we set it to 27, 20, 18 and 10; thus $\bS=diag(27,20,18,10)$. The entries in $\mathbf{E}^{(1)}$ and $\mathbf{E}^{(2)}$ are all generated as $i.i.d$ random samples from $N(0,\sigma^2)$, where $\sigma \in \{0.1,0.2,...,1\}$.

\subsection{Competing Methods}
We compare the performance of our method with four biclustering methods for multi-view data and one biclustering method with stability selection developed for data from one view. For the multi-view-based methods, we consider the proximal co-clustering \citep{multicoclustering} [mvProx], the multi-view svd [mvSVD] \citep{multisvd}, and the generalized biclustering (GBC) \citep{GBC} methods. We perform mvProx and mvSVD using the R-package \textit{mvcluster} (Version 1.0). This R package includes the mvSVD method as well as two proximal co-clustering methods using $l_1$-norm regularization [mvProxL1] and $l_0$-norm regularization [mvProxL0] respectively. The algorithms from the R package \textit{mvcluster} (mvSVD, mvProxL0 and mvProxL1)  detect one integrative bicluster at a time. To detect subsequent biclusters for the \textit{mvcluster} (mvSVD, mvProxL0 and mvProxL1) methods, we follow suggestions in  \cite{multicoclustering, multisvd}  and we manually subset the data after each run and we delete the samples (rows) that are detected previously. 
 We stack the views when applying the  biclustering with stability selection [s4vd] method, and we use the R-package \textit{s4vd} (Version 1.1.1). GBC is a Bayesian biclustering method for detecting biclusters from multiple views that allow each view to have a different probability distribution. Please refer to Section 1 in the supplementary material for description of these methods. The simulations have been carried out using the Minnesota Supercomputing Institute Mangi compute cluster. Simulations of mvSVD, mvProxL0, mvProxL1, and s4vd have been implemented with R 4.0.4, and simulations of iSSVD have been implemented with Python 3.7.
 
\subsection{Evaluation Criteria}
We evaluate the proposed and existing methods based on  bicluster similarity measures, F-score, and variable and sample selection. These are widely used measures in the statistical and machine learning literature for assessing biclustering methods \citep{Robust,Autodecoder}. For similarity measures, (i.e., similarity between the algorithm-generated biclusters and true biclusters from the same data)  we consider bicluster relevance and recovery, which are defined as follows. 

Suppose $\bM$ is the set of estimated biclusters and $\bM^*$ is the set of true biclusters, each containing a set of columns $\bC$ and a set of rows $\bR$. Let $\bM = \{\bM_1, \bM_2,...,\bM_m\}$ and $\bM^* = \{\bM^*_1, \bM^*_2, ..., \bM^*_n\}$ where $\bM_i = \bR_i \times \bC_i$ and $\bM^*_i = \bR^*_i \times \bC^*_i$, and $\times$ denotes the Cartesian product of the sets of rows and columns. Then the Jaccard index for two biclusters, each obtained from the Cartesian product is
$$Jac(M_i, M^*_i) = \dfrac{\bM_i \cap\bM^*_i}{\bM_i \cup \bM^*_i}.$$ Similarly as in  \cite{Robust}, the average relevance and recovery scores are defined as:
$$Relevance = \dfrac{1}{m} \sum_{i=1}^m \underset{j\in\{1,2,...,n\}}{\max} Jac(\bM_i, \bM^*_j)$$
$$Recovery = \dfrac{1}{n} \sum_{i=1}^n \underset{j\in\{1,2,...,m\}}{\max} Jac(\bM^*_i, \bM_j)$$
Essentially, the relevance score shows how well the detected biclusters represent the true ones  while the recovery score evaluates to what degree the true biclusters are recovered by the algorithm. 

For a combined effect of relevance and recovery measures, we consider the F-score, the harmonic mean of average relevance and recovery \citep{Autodecoder}: 
$$F-score = \dfrac{2\times Relevance \times Recovery}{Relevance + Recovery}. $$

For samples and variables selected, we consider  false positives (FP) and false negatives (FN). The FP is defined as the ratio of number of falsely selected non-zero elements outside of true bicluster against number of elements in the true bicluster. Conversely, the FN is defined as the ratio of number of non-zeros computed by the algorithm in the true bicluster against the number of elements in the true bicluster. 

\subsection{Simulation Results}
\subsubsection{Scenario One}

\begin{figure}[httb]
    \centering
    \includegraphics[scale=0.8]{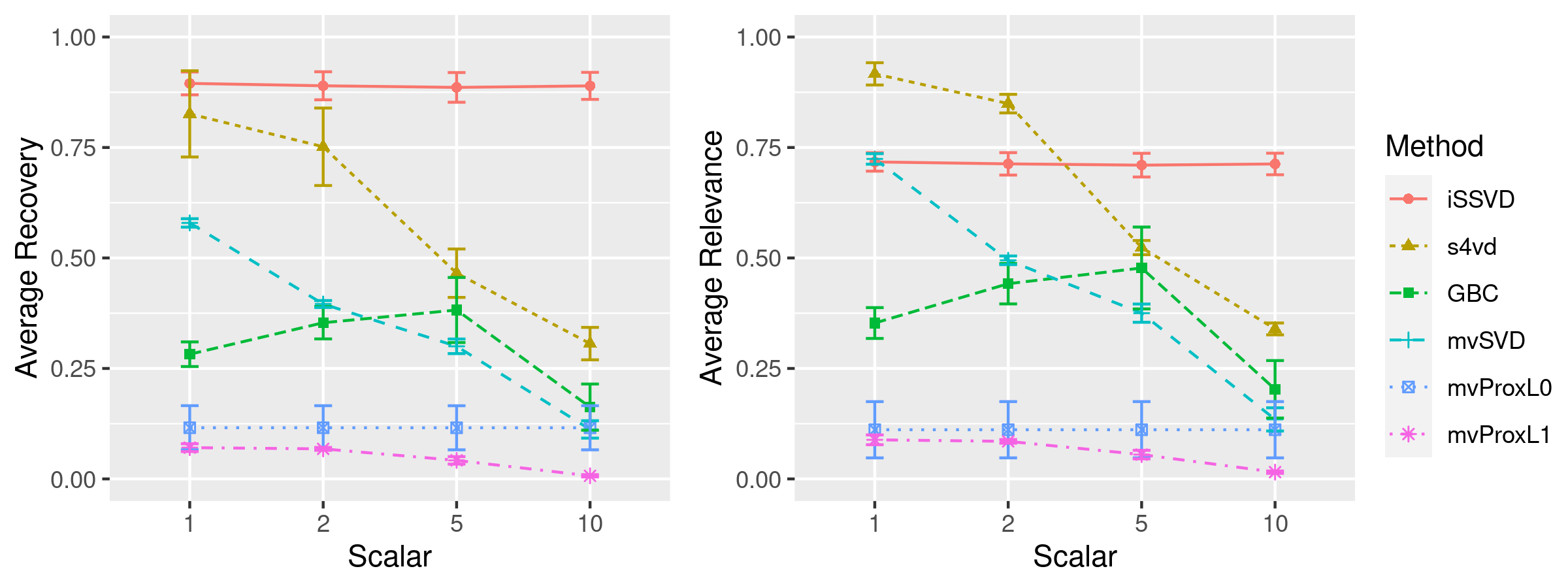}
    \caption{Simulation results for case 1. The boxplots show the distributions of average relevance and average recovery  over 50 randomly simulated multi-view data. Scalar indicates the scalar $s$ multiplied with the second view $\mathbf{X}^{(2)}$. 
    }
    \label{fig:scalar}
\end{figure}

~\\
\noindent\textit{Unbalanced scales}\\
In the first case of scenario one, we vary the scalar $s$ to be 1, 2, 5 and 10 to evaluate the methods in situations of unbalanced scales. The average recovery and relevance scores  are shown in Figure \ref{fig:scalar} for fixed noise level $\sigma=0.2$. 
For scalar $s=1$, i.e. both views on the same scale, iSSVD and s4vd had higher average recovery scores, whereas mvSVD and GBC had lower scores and mvProxL0 and mvProxL1 performed worse. 
The performance of iSSVD is slightly better and more stable than s4vd based on the scores of average recovery, since the median is higher and the variability is lower. From the relevance scores (Figure \ref{fig:scalar}, right panel), s4vd has a higher performance. This is because iSSVD sometimes identified more than 4 biclusters, and even though  it detected all true biclusters, it also assigned some noise to biclusters. 
As we increase the scales from  2 to 10, the two views become more and more unbalanced. We observe that the average recovery scores of iSSVD are still higher, suggesting that our method can perform well when data are unbalanced. However, for s4vd and mvSVD, their abilities to detect biclusters decrease dramatically; the performance of GBC increased for $s=2$ and $5$ but decreased for $s=10$. Furthermore, mvProxL0 and mvProxL1 perform poorly in all the situations. The average relevance scores of these methods show a similar trend but mvProxL0 tend to be better than mvProxL1. The relevance scores of iSSVD are suboptimal compared to recovery scores.  The relevance scores of s4vd are higher in less unbalanced scales ($s=1,2$) settings.  However, when the two views are more unbalanced, the biclusters detected by iSSVD are more representative of the true biclusters.   In the unbalanced scales, s4vd mostly is able to detect the biclusters from the more dominant view, which in this case is the second view. Supplementary Figures S4-S11 give results for Scenario 1, for  Case 1 and 2, when data are standardized, centered, or scaled. Compared to Figures 2 and 3, Figures S4-S11 suggest that centering as well as standardizing data so that each variable has mean zero and variance one prior to implementing the biclustering algorithms result in poor bicluster detection performance. Further, scaling each variable  to have variance 1 or each View to have Frobenius norm 1 and implementing iSSVD yield results that are comparable to the original data. 

~\\
\textit{Varying noise level}\\
The results for case 2 are shown in Figure \ref{fig:sigma} and Table S1.4 in the Supplementary Material. For noise level 0.1, the average recovery score for iSSVD is almost 1 whereas that of other methods is considerably lower. This indicates that iSSVD is able to detect all four integrative biclusters in the data. As the noise level increases so that the data become more corrupted, the performance of all methods deteriorate. When $\sigma$ is larger than 0.5, the median of average relevance of iSSVD is below 0.5. In noisy settings, the methods tend to detect more samples and variables outside the true biclusters as signals and propose them as biclusters, while  detecting the true biclusters less frequently. When $\sigma$ is larger than 0.8, the average relevance scores are almost close to 0, indicating that the abilities of these methods to detect biclusters that are representative of the true biclusters are largely impaired by noise. The average relevance scores show a  similar trend, but these are lower than the recovery scores; the methods start to assign
noise as other biclusters after detecting the true ones since we set the maximal number of biclusters to be detected as 5. In situations of noisy data, the algorithms tend to assign noise as biclusters and are less capable to detect true hidden structures.\\

\begin{figure}[httb]
    \centering
    \includegraphics[scale=.8]{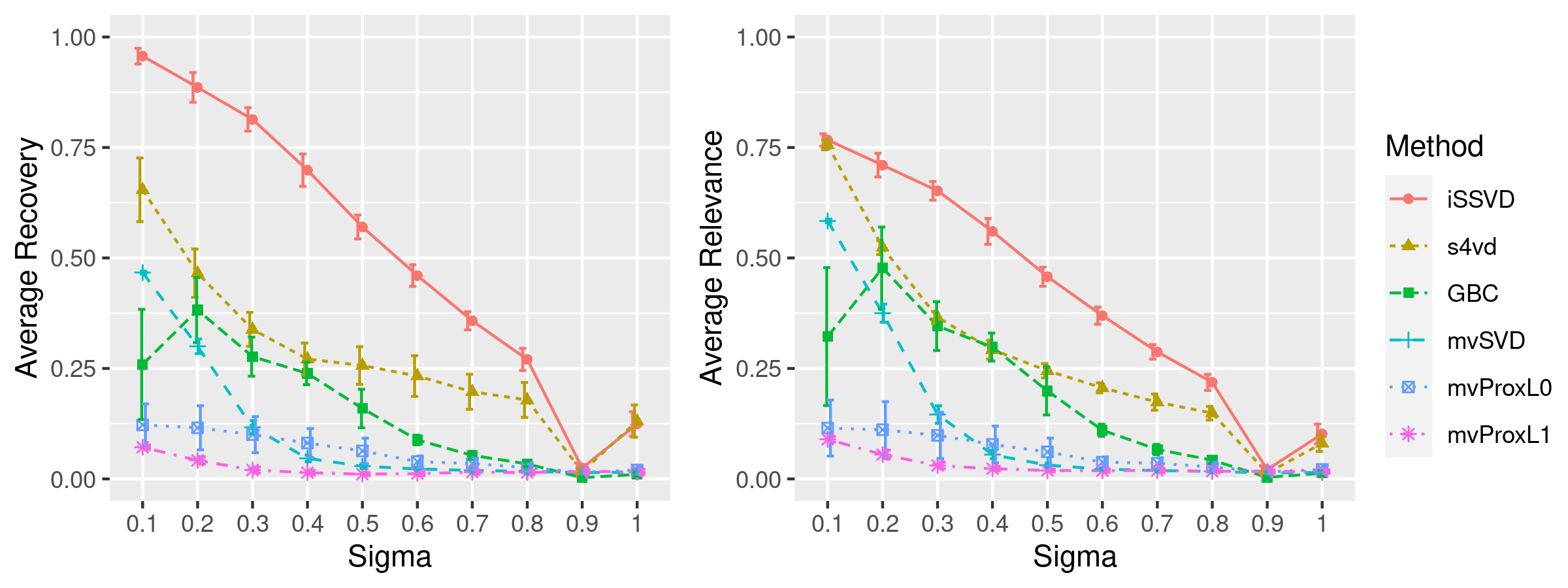}
    \caption{Simulation results for case 2. The boxplots show the distributions of average relevance and recovery over 50 randomly dimulated multi-view data. Sigmas indicate the value of noise level $\sigma$. Here we fix the scalar to be $s = 5$. The values of these indices and F-scores, FP and FN rates are described in supplementary material Table S1.4.}
    \label{fig:sigma}%
\end{figure}
\noindent\textit{Run times}\\
Next we show the runtimes measured in seconds for the biclustering algorithms in Supplementary Table S1.1. The time is captured from running every data set in Case 1, i.e. with dimensions $n=100$ and $p^{(1)}=p^{(2)}=1,000$, and averaged out to get the runtimes of each data set. The algorithms are carried out using the Minnesota Supercomputing Institute Mangi compute cluster. We use pointwise control for iSSVD and s4vd. The speed of iSSVD is comparable with mvProxL0, mvProL1 and mvSVD which are written in C++, but s4vd took approximately 10 times longer to finish running the same job. GBC took approximately 70 times longer to execute the same job.
In case 3, we use ultra-high dimensional setting to evaluate the computational efficiency of the algorithms. The dimensions of the data are $n=500$ and $p^{(1)}=p^{(2)}=10,000$. On average, iSSVD takes about 2 to 3 minutes for one Monte Carlo simulation while s4vd takes more than 10 minutes (Table S1.2). Furthermore, mvSVD, mvProxL0 and mvProxL1 all have difficulty to converge in this situation. GBC is computationally expensive and thus not practical to run for this case. 
~\\

\subsubsection{Scenario Two}

\begin{figure}
    \centering
    \includegraphics[scale=0.8]{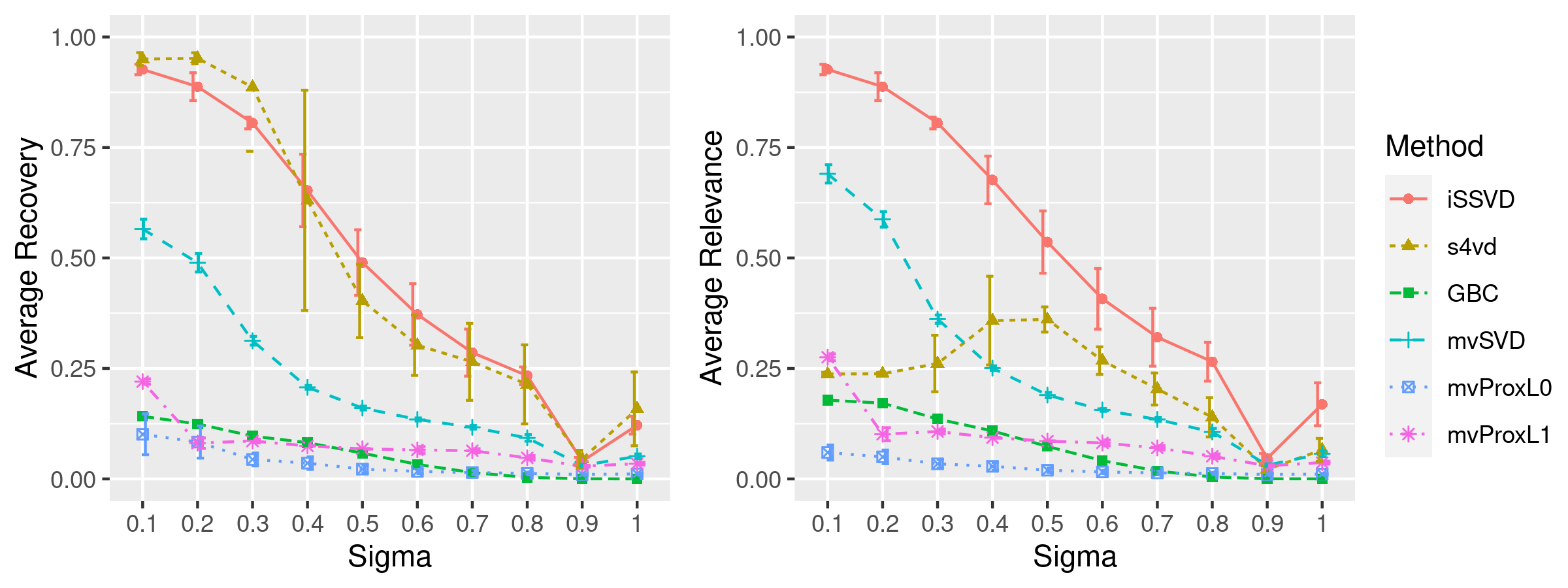}
    \caption{Simulation results for Scenario Two. The boxplots show the distributions of average relevance and recovery over 50 randomly simulated multi-view data. Sigmas indicate the value of noise level $\sigma$. Each randomly simulated data is of size $n=200$ and $p^{(1)}=p^{(2)}=1,000$. Each  bicluster in one view has 50 rows and 100 columns. We ensure that every sample (row) belongs to one of the four biclusters and that there are no overlaps. The average number of unclustered samples, F-scores, FP and FN rates are described in supplementary material.}
    \label{fig:scenario2}
\end{figure}

~\\
The simulation results of Scenario Two are shown in Figure \ref{fig:scenario2}. We set the maximum number of biclusters to be detected to 5, which is similar to Scenario 1. When the noise level $\sigma \leq 0.2$, iSSVD has higher relevance and recovery scores; the medians are almost 1. When $\sigma=0.1$, almost all samples are put into their right clusters by iSSVD and only approximately 8 samples are unclustered when $\sigma = 0.2$ (Table S2.1). For s4vd, the recovery scores are high but the relevance scores are significantly lower. This is because s4vd can only detect one integrative bicluster, remaining about three fourths of the samples unclustered. The performance of mvSVD is also worse than iSSVD. It leaves about one fourth of the samples unclustered and its scores for relevance and recovery are only around 0.5. Other multi-view methods could not detect meaningful integrative biclusters in this scenario, similarly to scenario one. As the noise level increases, s4vd has an unstable performance, as seen by the large variation in the boxplots. Meanwhile, the medians of iSSVD remain the highest among the five methods compared. We can still observe downward trend in the scores as $\sigma$ increases because the data become messier. We also note that the average relevance scores for s4vd increases, albeit lower than iSSVD,  but it has at least half of the samples unclustered over the range of $\sigma$ (Table S2.1). For iSSVD, about half of the samples are unclustered when $\sigma > 0.6$. Note that mvSVD can assign many samples to biclusters (as observed from the number of unclustered samples in Table S2.1) but it still achieves lower average relevance and recovery scores, since the samples are not correctly assigned  to the true sample clusters. 

Simulation results for Scenarios One and Two  suggest that the proposed method, iSSVD,  is better at detecting true biclusters, in both balanced and unbalanced scales settings, when compared to existing multi-view biclustering methods. Further, the proposed method is better for unbalanced scale settings when compared to the biclustering method with stability selection that is applicable to data from one view. 

\clearpage
\section{Real Data Analysis}
Chronic obstructive pulmonary disease (COPD) is a chronic progressive disease that affects at least 16 million adults in the US  and presents a substantial economic and social burden \citep{wheaton2015employment}. 
Like many complex diseases, COPD is characterized by a high degree of heterogeneity with many patients differing in their prognosis and response to therapy. Although the most common cause of COPD is cigarette smoking, not all smokers go on to develop COPD. Identifying subgroups of people  who may be at risk for developing COPD, and who are characterized by specific molecular features  is important to understand the heterogeneity in COPD and to improve outcomes for COPD patients. For instance, in \cite{chang2016copd},   gene expression data and a network-based clustering method was used to identify molecular subtypes for COPD. 
 
In this article, we carried out a study that integrated RNA sequencing and proteomics data from the COPDGene Study  to identify  COPD subgroups using the proposed biclustering method. The original data set has 1,317 proteins and 21,669  genes. There were 463 samples with proteomics, RNA sequencing and clinical/dempographic variables. Prior to applying the proposed and existing methods, we used univariate filtering to reduce the dimensionality of the data. In particular, we regressed each gene and protein with FEV$_{1}$ (percent predicted force expiratory volume in one second), and we retained genes and proteins with  p-value $< 0.05$ after adjustment for age, gender, and race. After filtering, we had 229 proteins and 4,415 genes for our analyses. The final datasets for the proteomics and RNA sequencing data were $\bX^{(1)} \in \Re^{463 \times 229}$ and $\bX^{(2)} \in \Re^{463\times 4,415}$, respectively.
 
We applied our algorithm and the other algorithms considered in the simulation section to the filtered data. \textit{s4vd} was applied to the concatenated filtered RNA sequencing and proteomics data.  For \textit{iSSVD} and \textit{s4vd}, we allowed the algorithms to detect the number of biclusters. We also did not allow for overlaps in rows and columns; i.e., each sample and variable is assigned to only one bicluster. This will allow us to identify distinct patient clusters that are characterized by distinct biomarkers. Four biclusters were detected by \textit{iSSVD} and 10 samples were unclustered. \textit{s4vd} did not detect any stable cluster. For \textit{mvProxL0, mvProxL1, and mvSVD}, we manually subset samples that are clustered when obtaining subsequent biclusters. In the end, 5 biclusters were obtained for mvProxL0, with 13 samples that were not clustered. \textit{mvSVD} and \textit{mvProxL1} detected 4 and 2 biclusters, respectively. One sample was not clustered for \textit{mvProxL1}. GBC identified 4 biclusters  but many samples identified by GBC belonged to multiple clusters. To avoid overlaps and to facilitate comparisons, we made random bicluster assignment for samples that overlapped. Please refer to the Supplementary Table S1.6 for our parameter settings. GBC identified 5 biclusters.

\noindent{\textbf{Clinical characteristics of  sample clusters identified}}\\
For all methods, we assigned the unclustered samples to the detected biclusters as follows. For each method, we obtained the first principal component (PC) from principal component analysis (PCA) of data for each sample cluster detected. That is, for each sample cluster, we used only the variables characterizing that sample to obtain the principal components. The first PC for a sample  cluster  explains  the maximum variation and can be used to summarize the information for that cluster. We correlated the first PC  for each sample cluster with  data for each sample that was not clustered. We then assigned that sample to the bicluster with the highest correlation. Thus, in the end, all samples belonged to one bicluster. The first iSSVD bicluster consisted of 109 samples, 213 genes, and 27 proteins. The second bicluster was made of 111 samples, 195 genes, and 25 proteins. The third bicluster had 127 samples, 165 genes, and 25 proteins. The fourth bicluster comprised of 116 samples, 169 genes, and 19 proteins.  

Compared to \textit{iSSVD}, all other  competing methods with the exception of \textit{mvProxL0}, did not find  meaningful sample subgroups from the data. The sample clusters detected by \textit{mvProxL1}, \textit{mvSVD} and \textit{GBC} were not differentiated on lung function (as measured by FEV$_1$/FVC, FEV$_1$ [\% predicted]), demographics/clinical variables, and symptoms (Tables S4.1 -  and S4.4 in the supplementary material). The sample clusters detected by \textit{mvProxL0} showed differences in some variables, but were not different across key lung function variables such as FEV$_1$/FVC and FEV$_1$ [\% predicted]). Compared to \textit{mvProxL0}, the sample clusters detected by our method showed differences in more variables.

The four sample clusters identified by  \textit{iSSVD} where well-differentiated on some key demographic, clinical, and outcome variables such as age, FEV$_{1}$/FVC ratio, and BODE index (Table \ref{tab:Table1}).  The FEV$_{1}$/FVC ratio is widely used to diagnose COPD. FEV$_{1}$ (forced expiratory volume in one second) is the volume of breath exhaled in one second, and it is used to gauge severity of COPD. The BODE (body mass index, obstruction, dyspnea, and excercise) index is a multidimensional assessment of an individual's risk of death \citep{celli2004body}; higher values suggest increased risk. From Table \ref{tab:Table1}, individuals in Biclusters 1, 3, and 4 have lower lung function as depicted by FEV$_{1}$  compared to individuals in Biluster 2. Bicluster 1 has a lower mean FEV$_{1}$/FVC ratio and lower mean FEV$_{1}$ value, followed by Biclusters 3 and 4. Individuals in Bicluster 2 have better lung function as can be observed by the higher FEV$_{1}$/FVC ratio, FEV$_{1}$ percent predicted and lower mean dyspnea and BODE index values. Using the terminology in  \cite{chang2016copd}, we refer to Bicluster 1 as the ``severely affected" group, Bicluster 3 as the ``moderately affected" group, and ``Bicluster 4" as the less affected group. Also, we refer to Bicluster 2 as the ``preserved lung function" group since the mean FEV$_{1}$/FVC ratio and FEV$_{1}$ values were on average within normal limits. Bicluster 1, the severely affected group, also had a higher BODE index and a  higher dyspnea score. They were more likely to be old, males and they tended to be previous smokers.  Bicluster 3, the moderately affected group, had the next highest dypsnea score, BODE index, and frequency of exacerbation, followed by Bicluster 4, the less affected group. The preserved lung function group, Bicluster 2,  was predominantly females, who were less likely to have COPD as defined by FEV$_{1}$/FVC ratio, reported less symptoms (e.g., lower emphysema as measured by Thirona and Percent 15), had lower mean age, and tended to be current smokers. 

Five-year follow-up data were available for $162$ individuals. We observe that the subgroups identified by iSSVD are again well-differentiated on some key outcomes, clinical variables and symptoms (Table S4.5). Biclusters 1, 3, and 4 again had lower FEV$_1$/FVC, lower FEV$_1$ percent predicted, higher BODE index and  higher dyspnea score. The  group with preserved lung function (Bicluster 2) had lower dyspnea score, lower BODE index, better lung function, and were less likely to have visited the emergency room (ER) or to be hospitalized for lung problems. 

\noindent\textbf{Subtype-specific biologic pathway and disease enrichment analysis for variable clusters characterizing sample clusters identified by \textit{iSSVD}}

We used the Ingenuity Pathway Analysis (IPA) software  to investigate the molecular and cellular functions, pathways, and diseases  enriched in the proteins and genes identified for each bicluster. IPA searches the ingenuity pathway knowledge base, which is manually curated from scientific literature and over 30 databases, for gene interaction. We focused on the variable clusters identified by \textit{iSSVD} since the sample subgroups identified by this method were well-differentiated on many clinical, demographic, and outcome variables. We observed strong enrichment of functional pathways (Supplementary Material Tables S4.6-S4.7). Some of the significantly enriched canonical pathways that mapped to the gene and protein lists for the severely affected group (Bicluster 1) included IL-8 signaling, mTOR pathway, and  Intrinsic Prothrombin activation pathway. The mTOR signaling pathway is involved in many cellular processes such as cell growth, metabolism, and survival \citep{jewell2013nutrient}.  Research suggests that the activation of the  mTOR signaling pathway can induce cell senescence in the lung, which in turn can result in COPD \citep{houssaini2018mtor}. 

Some pathways enriched for the moderately affected group (Bicluster 3) included airway pathology in COPD and oxidative phosphorylation.  Also, the inflammasome pathway,  airway pathology in COPD, and IL-17 signaling pathways were over-represented in our gene and protein lists for the less affected group (Bicluster 4). Some pathways enriched for the  preserved lung function group (Bicluster 2) included taurine biosynthesis and enhanced cardiac hypertrophy signaling pathway. While there were overlaps in some of the enriched pathways for the clusters, there were also unique pathways identified.  


In addition to IPA canonical pathways, proteins and genes were also categorized to related diseases and functions. Again, there were some overlaps in  the top 5 enriched diseases and functions for the clusters (Tables S4.11 - S4.12). The severely affected group (Bicluster 1) was characterized by diseases  that included cardiovascular (such as atherosclerosis, ventricular dysfunction and peripheral vascular disease) and inflammatory (such as chronic inflammatory disorder) diseases. Some of the diseases characterizing the moderately affected group (Bicluster 3) included cardiovascular (such as infarction and ischemia of brain) and cancer. The less affected group (Bicluster 4) was characterized by neurological (including abnormal regeneration by peripheral nervous system and cerebrovascular dysfunction) and hereditary disorder. Also, the preserved lung function group (Bicluster 2) was characterized by inflammatory response (such as inflammation of lung) and cancer.  We also used IPA for network analysis to connect key genes, proteins, and enriched categories of diseases and functions. Our results showed that the severely affected group was characterized by the cardiovascular disease, organismal injury and abnormalities, hematological system development and function (based on our protein list). The  preserved lung function group was characterized by the dermatological diseases and conditions, organismal injury and abnormalities, organismal development network (from our protein list). 


\begin{table}[ht]
\centering
\begin{scriptsize}
\resizebox{\textwidth}{!}{%
\begin{tabular}{lccccc}
\hline
\hline
Bicluster (N) & 1 (109) & 2 (111) & 3(127) & 4 (116) & P-value \\
\hline
\hline
\textbf{Demographics and Clinical} & \\
Age (years) & 71.73 (7.00)	&63.28 (8.21)	&66.80 (8.47)&	67.85 (8.04) &  \textless{}.00001 \\
Body Mass Index & 29.46 (5.80)	&29.17 (6.01)&	28.34 (6.56)	&29.77 (6.55) &  0.2333 \\
Pack Years & 50.77 (26.53)	&39.57 (19.85)&	43.94 (24.85)	&46.46 (26.69) &  0.01433 \\
Duration of smoking (years) & 37.82 (12.01)	&36.22 (12.11)	&36.60 (11.58)	&36.76 (12.36) & 0.8997 \\
Gender & &\\
~~~~Males & 65 (60\%)	&52 (47\%)	&58 (46\%)	&60 (52\%)&	 0.1402 \\
~~~~Females &44 (40\%)	&59 (53\%)	&69 (54\%)	&56 (48\%)&	 \\
Smoking Status & &\\
~~~~Former & 94 (86\%)&	66 (59\%)	&98 (77\%)&	94 (81\%)&	\textless{}0.0001 \\
~~~~Current&  15 (14\%)	&45 (41\%)&	29 (23\%)&	22 (19\%)&	  \\
Dyspnea Score (MMRC)  & 1.34 (1.45) & 0.77 (1.17) & 1.18 (1.28) & 1.02 (1.27) &  0.0048 \\
BODE Index & 2.20 (2.44) & 0.90 (1.51) & 1.48 (2.07) & 1.18 (1.64) &  0.0005\\
\textbf{Outcomes} & &\\
FEV$_{1}$/FVC & 0.60 (0.17)&	0.72(0.11)	&0.67(0.16)	&0.66(0.15) &  \textless{}0.00001 \\
FEV$_1$ (\% predicted) & 70.43(27.88)	&85.23 (21.04)&	78.63 (25.75)	&77.07 (24.51) &  0.0026 \\
COPD status& & &\\
~~~~No	&35 (34\%)	&56 (57\%)	&57 (51\%)	&48 (47\%)&	0.0096\\
~~~~Yes	&67 (66\%)	&42 (43\%)	&54 (49\%)&	54 (53\%)&	\\
\textbf{Symptoms} & & \\
Exacerbation Frequency & 0.29 (0.86) & 0.19 (0.56) & 0.20 (0.58) & 0.16 (0.49) & 0.6571 \\
Percent Emphysema (Thirona) & 9.66 (11.19)	&3.45 (5.25)	&8.52 (12.76)	&7.26 (10.52) & \textless{}0.0001\\
Percent 15 & -931.71 (25.00)&	-919.20 (19.48)&	-926.60 (28.42)&	-925.15 (25.57) & 0.0021\\
Ever had Asthma & &\\
~~~~No & 104 (95.4\%)&	109 (98.2\%)&	125 (98.4\%)&	112 (96.6\%)&	 0.4798 \\
~~~~Yes & 5 (4.6\%)&	2 (1.8\%)&	2 (1.6\%)&	4 (3.4\%)&	 \\
 Gastroesophageal Reflux & &\\
~~~~No & 69 (63\%) & 86 (77\%) & 79 (62\%) & 74 (64\%) &  0.0473 \\
 ~~~~Yes & 40 (37\%)&	25 (23\%)&	48 (38\%)&	42 (36\%)&	 \\
Been to ER or hospitalised for lung problems & &\\
~~~~No & 98 (90\%) & 106 (95.5\%) & 116 (91.3\%) & 108 (93.1\%) &  0.4274 \\
~~~~Yes & 11 (10\%)	&5 (4.5\%)&	11 (8.7\%)	&8 (6.9\%) &   \\
\hline
\hline
\end{tabular}%
}
\caption{Clinical characteristics of iSSVD biclusters of COPD data. Values are mean (SD) for continuous variables, and N (percentages) for binary/categorical variables. P-values are from comparing all four groups using Kruskal-Wallis test for continuous variables, and Chi-square test for categorical variables. MMRC- modified Medical Research Council dyspnea score (0-4, 4= most severe symptoms). GOLD stages 1-4 are combined to form COPD cases; GOLD stage 0 form COPD controls.}\label{tab:Table1}
\end{scriptsize}
\end{table}

\section{Discussion and Conclusion}
In this manuscript, we extended existing biclustering method based on sparse singular value decomposition for data from one view \citep{Lee} to data from multiple views.  Our method followed ideas in \cite{Robust} and incorporated stability selection \citep{meinshausen2010stability}, a subsampling based variable selection that allows to control Type I error rates. The proposed algorithm estimates the probability of samples and variables to belong to a bicluster, finds stable biclusters, and results in interpretable row-column associations. The proposed algorithm, developed in Python 3, is computationally efficient and user-friendly and will be useful in many disease subtyping applications. Through simulation studies, we showed that the proposed method outperforms several other single- and multi-view biclustering methods in detecting artificial biclusters.

When our method was applied to RNA sequencing and proteomics data from the COPDGene study, we detected four biclusters that were well-differentiated by some demographics and  clinical variables as well as key COPD outcomes.   Three  biclusters which we call severely, moderately, and less affected groups, seemed to have poor lung function and clinical outcomes, while one bicluster, which we call preserved lung function group seemed to have better (preserved) lung function and clinical outcomes. 
We also performed enrichment analysis of the genes and proteins characterizing the sample clusters. While certain biological processes were most enriched in specific biclusters, there was also notable overlap in processes across biclusters. Particularly enriched molecular and cellular functions included cellular movement, cellular growth and proliferation and cell death and survival. 

The limitations of our work include the following. First, when applied to real data sets, a smaller preset Type-I error rate tends to yield small biclusters, which results in more unclustered samples. However, increasing the error rate might compromise the strength against noise.  Second, we mention  ways to choose the number of biclusters $K$ but these could be improved, especially in situations of noisy data. Third, we noticed that under regular settings, the run times of iSSVD and s4vd tend to be considerably longer than point-wise control settings. However, since we are not searching for the entire range of regularization parameters in point-wise control, it is likely that we overlook the optimal regularization parameters. Thus, future work can seek to  improve the computational time to accommodate the computational demands for stability selection  that searches the entire range of regularization parameters.  

In conclusion, we have developed a biclustering method for multi-view data capable of detecting stable row and column clusters.   The encouraging simulation and real data findings  motivate further applications to identify disease subtypes and subtype-specific molecular features.  

\section*{Funding and Acknowledgements}
The project described was supported by grant 5KL2TR002492-03 from the National Institutes of Health and by Award Number 1R35GM142695-01 from the National Institute Of General Medical Sciences, Award Number U01 HL089897 and Award Number U01 HL089856 from the National Heart, Lung, and Blood Institute. COPDGene is also supported by the COPD Foundation through contributions made to an Industry Advisory Board that has included AstraZeneca, Bayer Pharmaceuticals, Boehringer-
Ingelheim, Genentech, GlaxoSmithKline, Novartis, Pfizer, and Sunovion.

Disclaimer: The views expressed in this article are those of the authors and do not reflect the views of the United States Government, the Department of Veterans Affairs, the funders, the sponsors, or any of the authors’ affiliated academic institutions.


~\\
\textit{Declaration of Conflicting Interests}: The Authors declare that there is no conflict of interest.

\section*{Data Availability}
The data used were provided by the COPDGene Study group. COPDGene clinical and RNASeq data are available on dbGap.  
We provide a Python package, \textit{iSSVD}, to facilitate the use of our method. Its source  codes, along with a README file are available via this link: \url{https://github.com/weijie25/iSSVD}.

\section*{Supplementary Data}
In the online Supplementary Materials, we provide more results from simulations and real data analyses.

\end{document}


\title{Supplementary material for  Robust Integrative Biclustering for Multi-view Data }
	\begin{tiny}
		\author{W. Zhang$^{\text{\sf 1}}$, C. Wendt$^{\text{\sf 2}}$, R. Bowler$^{\text{\sf 3}}$, C.  P. Hersh$^{\text{\sf 4}}$, and S. E. Safo$^{\text{\sf 1}}$\thanks{corresponding author email: ssafo@umn.edu}\\[4pt]
			$^{\text{\sf 1}}$Division of Biostatistics, $^{\text{\sf 2}}$Division of Pulmonary, Allergy and Critical Care \\
			University of Minnesota, Minneapolis, 55455, USA \\
			$^{\text{\sf 3}}$Division of Pulmonary, Critical Care and Sleep Medicine, \\ Department of Medicine, National Jewish Health, Denver, CO, USA  \\
			$^{\text{\sf 4}}$Channing Division of Network Medicine, Brigham and Women's Hospital\\ Harvard Medical School, Boston, MA, USA}
	\end{tiny}
	\markboth%
	{Zhang et.al.}
	{Robust Integrative Biclustering}
	
	\maketitle
	\def\spacingset#1{\renewcommand{\baselinestretch}%
		{#1}\small\normalsize} 
	\maketitle
	
	\spacingset{1}

\section{Brief overview of existing multi-view biclustering methods considered} 

A multi-view biclustering method groups samples (rows) across multiple views and identify variables (columns) within each view simultaneously.
Several multi-view biclustering (co-clustering) methods have been proposed over the past few years. The R package `mvcluster' contains three algorithms mvProxL0 (mvlrrl0, \citep{multicoclustering}), mvProxL1 (mvlrrl1, \citep{multicoclustering}) and mvSVD (mvsvdl1, \citep{multisvd}). Here mvProxL0 is based on sparse low-rank matrix approximation where the decomposed components are regularized by the $l_0$-norm. It solves the following optimization problem:
\begin{align*}
 \nonumber\underset{\bomega,\bu^k,\bv^k,k=1,...,m}{\min} \sum_{k=1}^m \|\bX^k - diag(\bomega)\bu^k\bv^{k^T}\|^2_F \\
 \text{subject to} \: \|\bomega\|_0 \leq s_{\bomega}, \|\bv^k\|_0 \leq s_{v^k} \\,
 \nonumber k=1, ..., m, \\
 \nonumber \bomega \in \mathcal{B}_n,
\end{align*}
where $\bomega$ is a binary vector of size $n$ that serves as a common factor to link multiple views, and for an arbitrary vector $\bz$, the $l_0$-norm $\| \bz\|_0$ counts the total number of nonzero elements of $\bz$. When $\omega_i = 0$, regardless of the values of the ith component of all $\bu^k$'s, the ith row will be excluded from the bicluster. $\mathcal{B}_n$ is the set that contains all binary vector of size $n$. $s_{\bomega}$ and $s_{v^k}$'s are hyper parameters to enforce sparsity of $\bomega$ and $\bv^k$'s. The optimization problem of mvProxL1 is similar except that it regularizes decomposed components by $l_1$ norms.

\newpage Our proposed method, integrative sparse singular value decomposition (iSSVD), is similar to mvSVD; it is based on singular value decomposition where the left singular vectors are used to identify rows in biclusters and the right singular vectors are used to identify columns. mvSVD extends single-view biclustering method SSVD to multi-view problems. The optimization objective, for example, when there are two views $M_1$ and $M_2$, is to solve:
\begin{align*}
 \nonumber \underset{\bz,\sigma_i,\bu_i,\bv_i,i=1,2}{\min} \|\bM_1-\sigma_1(\bz \odot \bu_1) \bv_1^T\|_F^2 + \|\bM_2 - \sigma_2(\bz \odot \bu_2)\bv_2^T\|_F^2  \\
 \nonumber + \lambda_z\|\bz\|_0 + \lambda_{v_1} \|\sigma_1 \bv_1\|_0 + \lambda_{v_2}\|\sigma_2\bv_2\|_0, \\
 \text{subject to} \|\bu_i\|_2 = 1, \|\bv_i\|_2 = 1, i = 1,2, \bz \in \mathcal{B}_n,
\end{align*}
where $\bz$ is a binary vector of size $n$ that serves as a common factor to link the views. Each component of $\bu_i, i=1,2$ is multiplied by the corresponding element in $\bz$, i.e. $\bu = \bu \odot \bz$. Thus, the authors enforce sparsity of $\bz$ and this will further induce sparsity of $\bu_1$ and $\bu_2$. $\lambda_z, \lambda_{v_1}$ and $\lambda_{v_2}$ are tuning parameters that balance approximation errors and regularization terms. The corresponding algorithms in the R package were compared with our method in the simulation section. Other than mvSVD which picked up some structures in the data, mvProxL0 and mvProxL1 performed poorly. Furthermore, these algorithms are not user friendly for several reasons. Firstly, the all algorithms can only detect one bicluster at a time. The user will have to manually subset the data if more than one bicluster is desired. Secondly, the parameter specification can cause extra unnecessary work. For example, the package only mentions the input parameter 'lz' of mvSVD is used to control the sparsity of $\bz$ and larger value will yield more sparsity. However, the user will have to perform some tuning to search for the optima function inputs. Several other parameters also have the same problem with no default values.

\cite{GBC} proposed a generalized biclustering (GBC) method for data from multiple sources. GBC assumes that the conditional distribution of the data given the latent components $\mathbf{U}$ and $\mathbf{V}$ (corresponding to  $\mathbf{Z}$ and $\mathbf{W}$ respectively in their method) comes from exponential family, and imposes the exponential family likelihood. To obtain the biclusters, the authors impose  Bayesian Laplacian shrinkage priors on $\mathbf{U}$ and $\mathbf{V}$, and then estimate these components from their likelihoods using an EM algorithm. Once $\mathbf{U}$ and $\mathbf{V}$ have been estimated, the product of the $k$th row of $\mathbf{U}$ and $k$th column of $\mathbf{W}$ form the $k$th bicluster. 

\section{More on stability selection} 
If all penalty parameters are the same, this will then correspond to stacking the data and applying the robust biclustering method proposed in \cite{Robust}. We note in the last paragraph on page 10 that this is likely to result in a too large or too small a tuning parameter for a particular view, which may in turn result in a solution that is trivial or not sparse.    This is true because ${\widetilde{\bv}}_{1}$ in equation (11) in the main text reduces to:
   \begin{equation}\label{eqn:compv}
       \widehat{\widetilde{v}}_{1j} = \text{sign}\{(\bX^{\smt} \bu_1)_j\} (|(\bX^T \bu_1)_j| - \lambda_{\min_{v_1}}/2)_{+},  
   \end{equation}
 where now ${\widetilde{\bv}}_{1} \in \Re^{p^{(1)} + \ldots + p^{(D)}}$, and $\bX$ is the stacked data with dimension $n \times (p^(1) + \ldots + p^(D))$. Then $\lambda_{\min_{v_1}}$ will correspond to the tuning parameter in the regularization region that ensures that the average number of selected variables in a bicluster is bounded by $e_{\Lambda_{v_1}}$, i.e.,  $q_{\Lambda_{v_1^(d)}} \le e_{\Lambda_{v_1}}$, where $e_{\Lambda_{v_1}} = \sqrt{E({\bv_1})(2\pi_{thr} -1)(p^{(1)} +\ldots+p^{(D))}}$. To put things into perspective, suppose we have two views, $n=100$, $p^(1)=150$, $p^(2)=200$. If we assume a Type 1 error of 0.05 in selecting variables (i.e., we expect about $0.05 \times 350=17.5$ variables (out of 350) to be false, i.e, $E(\mathbf{v}_1) =17.5$, then we choose a $\lambda_{\min_{v_1}}$ that will ensure that the average number of variables selected in a bucluster is bounded by $\sqrt{17.5(2\times0.8-1)(150+200)}=60.63$, assuming a selection threshold $\pi_{thr}=0.8$. Assuming this bound, it is likely that all $60.63$ variables will be chosen from the first $p^{(1)}$ or $p^{(2)}$ variables; if that was the case, this will not correspond to a 0.05 error rate for either view, but the error rate will be inflated (i.e., $30.3$ for $p^{(2)}$ and $40.4$ for $p^{(1)}$). Thus, when there are 2 or more views, it is ideal to consider an integrative approach that will not assume equal regularization parameters for all views, but will estimate stable biclusters for each view while controlling for view-specific Type 1 error

\section{More simulation results}
We provide some more simulation results in this section. Tables \ref{TableS11} and \ref{TableS15} give the run times for different methods in Scenario 1 case 1 and case 3 respectively. mvSVD, mvProxL0 had shorter run times in case 1 but in case 3 where the dimension of the two views were very high, these methods did not converge. Also, the run times for the proposed method iSSVD is considerably lower than s4vd on concatenated data. GBC is computationally expensize and thus impractical to run in the ultra-high dimensional case. 

Table \ref{TableS12} gives simulation results for Scenario 1 case 1 under varying scalar levels for unstandardized data. Here, the noise level $\sigma$ is fixed at $0.2$. We note that the for large scalar values (Scalar $>$2), the proposed method had better estimates compared to s4vd on concatenated data. The performance of the multi-view biclustering methods compared are suboptimal. 

Table \ref{TableS13} gives simulation results for Scenario 1 case 2 for fixed scalar level (scalar fixed at 5) and varying noise levels. We note that the performance of all methods compared deteriorate  when the noise level in each view increases. Nevertheless, the performance of the proposed method is better than the other methods. 

Table \ref{TableS21} gives the number of samples that remain unclustered in Scenario 2 for varying noise levels. We observe that more samples become unclustered by the proposed method as the noise level increases. sv4d and mvProxL0 had more unclustered samples even when the noise level was low. mvSVD and mvProxL1 had less unclustered samples but the relevance, recovery, and F-scores for these methods are lower which suggests that they are not detecting the true clusters.

We consider results when we center, standardize or scale each variable in each view and when we scale each view, too and we compare results to the results from the original data. Figures S4-S11 show that the proposed methods applied to centered or standardized  view still result in better performance than s4vd and the multi-view biclustering methods even for unbalanced scales. These findings indicate that standardizing each view and stacking the data and then applying biclustering with stability result in suboptimal performance. Also, multi-view biclustering method with stability selection is better than the existing multi-view biclustering methods without stability selection. Compared to Figures 2 and 3 in the main text, Figures S4-S11 suggest that centering as well as standardizing data so that each variable has mean zero and variance 1 prior to implementing the biclustering algorithms result in poor bicluster detection performance. Further, scaling each variable  to have variance 1 or each view to have Frobenius norm 1 give results that are comparable to the original data.

Table S2.3 shows the number of biclusters detected by the proposed method for scenario 2 using criteria a) in Remark 2. When the noise level was set at 0.1, we detected the true number of biclusters all the time out of the 50 Monte Carlo datasets. The proportion of the true number of biclusters detected decreased with increasing noise level. 

Table S1.5 gives the parameter settings used in the simulations. 

\begin{figure}
    \centering
    \renewcommand{\figurename}{Figure S}
    \hspace*{-1.5cm}
    \includegraphics[scale=0.6]{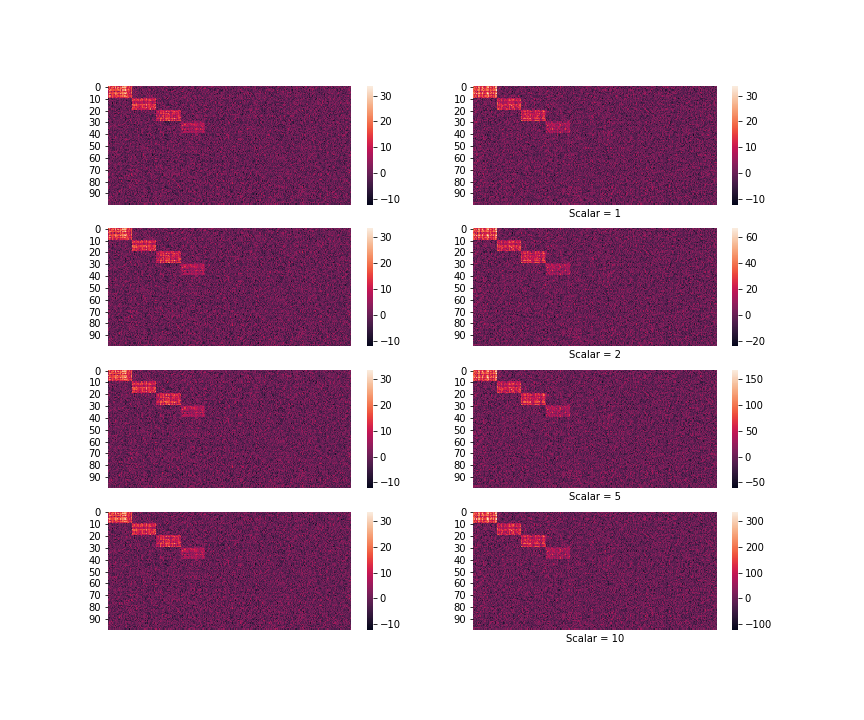}
    \caption{Data set-up of scenario one case 1.}
    \label{fig:s1}
\end{figure}

\begin{figure}
    \centering
    \renewcommand{\figurename}{Figure S}
    \hspace*{-1.5cm}
    \includegraphics[scale=0.6]{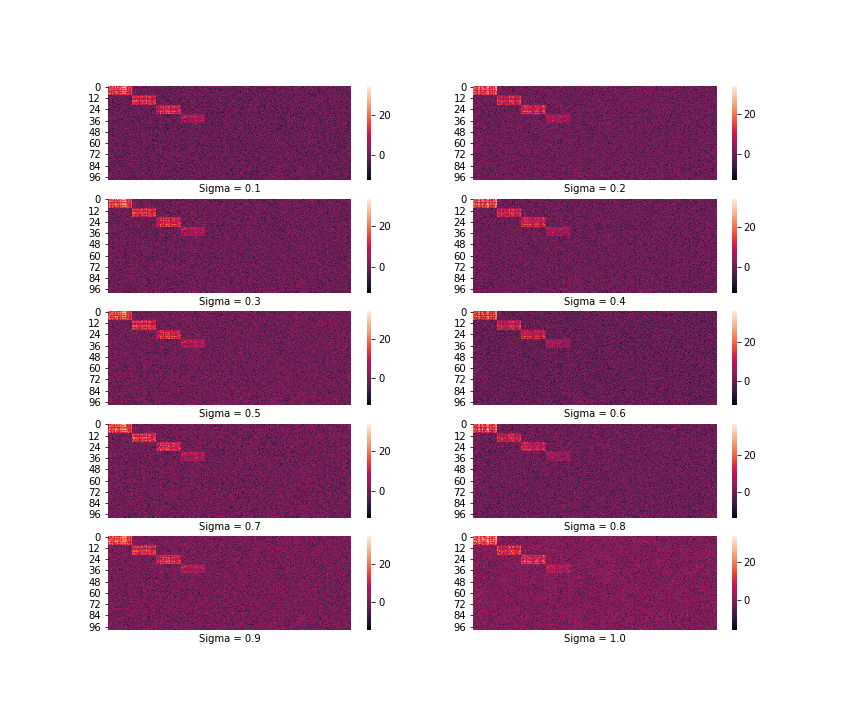}
    \caption{Data set-up of scenario one case 2.}
    \label{fig:s1}
\end{figure}

\begin{figure}
\renewcommand{\figurename}{Figure S}
    \centering
    \hspace*{-1.5cm}
    \includegraphics[scale=0.6]{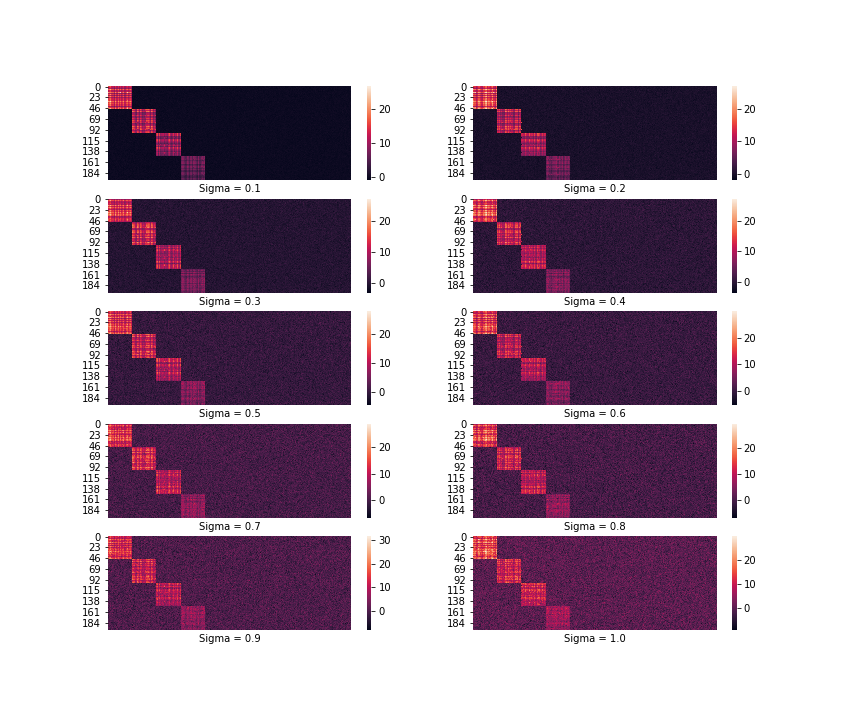}
    \caption{Data set-up of scenario two.}
    \label{fig:s2}
\end{figure}

\begin{table}[ht!]
\renewcommand\thetable{S1.1} 
\centering
\caption{Run time of different algorithms. The time is captured with all samples ran in scenario 1. The values are mean (SD). iSSVD and s4vd algorithms are run using pointwise control.}
\label{TableS11}
\begin{tabular}{cc}
\hline
\textbf{Algorithms} & \textbf{Time (s)} \\ \hline
iSSVD & 1.26 (0.43) \\
s4vd & 10.2 (1.25) \\
mvSVD & 0.421 (0.057) \\
mvProxL0 & 0.861 (0.010) \\
mvProxL1 & 0.276 (0.046) \\
GBC & 116 (19.656) \\ \hline
\hline
\end{tabular}
\end{table}

\begin{table}[ht!] 
\renewcommand\thetable{S1.2}
\centering
\caption{Run time of different algorithms in scenario 1 case 3. The values are mean (SD).  iSSVD and s4vd algorithms are run using pointwise control.The algorithms from R package ‘mvcluster’ did not converge in this situation. GBC is computationally expensive making it impractical to run for the dimensions considered.}
\label{TableS15}
\begin{tabular}{ll}
\hline
\textbf{Algorithms} & \textbf{Time (s)} \\ \hline
iSSVD & 140 (28.7) \\
s4vd & 1237 (531) \\
mvSVD & N/A \\
mvProxL0 & N/A \\
mvProxL1 & N/A \\
GBC & N/A \\ \hline
\end{tabular}
\end{table}

\begin{table}[]
\renewcommand\thetable{S1.3}
\centering
\caption{Simulation results of scenario 1 case 1. Noise level $\sigma$ is fixed at 0.2. The values are mean (SD).}
\label{TableS12}
\resizebox{\textwidth}{!}{%
\begin{tabular}{ccccccc}
\hline
\textbf{} & \textbf{Scalar} & \textbf{Ave. Recovery} & \textbf{Ave. Relevance} & \textbf{F-score} & \textbf{FP Rate} & \textbf{FN Rate} \\ \hline
\textbf{iSSVD} & 1 & 0.895 (0.0259) & 0.717 (0.0209) & 0.796 (0.0231) & 0.183 (0.0431) & 0.223 (0.00446) \\
 & 2 & 0.89 (0.0318) & 0.713 (0.0254) & 0.792 (0.0282) & 0.187 (0.0476) & 0.223 (0.00419) \\
 & 5 & 0.886 (0.0336) & 0.710 (0.0268) & 0.788 (0.0298) & 0.180 (0.0439) & 0.223 (0.00434) \\
 & 10 & 0.889 (0.0306) & 0.713 (0.0244) & 0.791 (0.0271) & 0.174 (0.0405) & 0.224 (0.00469) \\ \hline
\textbf{s4vd} & 1 & 0.826 (0.0976) & 0.917 (0.0252) & 0.866 (0.0591) & 0.0595 (0.0308) & 0.0319 (0.00625) \\
 & 2 & 0.752 (0.0877) & 0.849 (0.0211) & 0.795 (0.0522) & 0.0949 (0.0238) & 0.0704 (0.00821) \\
 & 5 & 0.466 (0.0548) & 0.523 (0.0160) & 0.491 (0.0334) & 0.167 (0.0223) & 0.356 (0.0153 ) \\
 & 10 & 0.306 (0.0367) & 0.339 (0.0134) & 0.321 (0.0234) & 0.236 (0.0303) & 0.504 (0.00211) \\ \hline
\textbf{mvSVD} & 1 & 0.579 (0.00956) & 0.724 (0.0119) & 0.644 (0.0106) & 0.207 (0.0659) & 0.00854 (0.00436) \\
 & 2 & 0.396 (0.00784) & 0.495 (0.00981) & 0.440 (0.00872) & 1.58  (0.0473) & 0.00418 (0.00188) \\
 & 5 & 0.300 (0.0165) & 0.375 (0.0206) & 0.333 (0.0183) & 3.57 (0.130) & 0.0156 (0.0172) \\
 & 10 & 0.112 (0.0197) & 0.135 (0.0263) & 0.122 (0.0222) & 4.79 (2.25) & 0.265 (0.0735) \\ \hline
\textbf{mvProxL0} & 1 & 0.116 (0.0499) & 0.111 (0.0636) & 0.113 (0.0565) & 1.11 (0.0872) & 0.794 (0.0872) \\
 & 2 & 0.116 (0.0499) & 0.111 (0.0636) & 0.113 (0.0565) & 1.11 (0.0872) & 0.794 (0.0872) \\
 & 5 & 0.116 (0.0499) & 0.111 (0.0636) & 0.113 (0.0565) & 1.11 (0.0872) & 0.794 (0.0872) \\
 & 10 & 0.116 (0.0499) & 0.111 (0.0636) & 0.113 (0.0565) & 1.11 (0.0872) & 0.794 (0.0872) \\ \hline
\textbf{mvProxL1} & 1 & 0.0708 (0.00885) & 0.0885 (0.0111) & 0.0787 (0.00983) & 0 (0) & 0.285 (0.0876) \\
 & 2 & 0.0680 (0.00331) & 0.0850 (0.00414) & 0.0756 (0.00368) & 0 (0) & 0.255 (0.0354) \\
 & 5 & 0.0419 (0.00843) & 0.0551 (0.00969) & 0.0476 (0.00905) & 0 (0) & 0.317 (0.0842) \\
 & 10 & 0.00704 (0.00262) & 0.0153 (0.00267) & 0.00954 (0.00291) & 0 (0) & 0.422 (0.0562) \\ \hline
\textbf{GBC} & 1 & 0.282 (0.028) & 0.353 (0.035) & 0.314 (0.031) & 0.0 (0.0) & 0.06 (0.011) \\
& 2 & 0.353 (0.037) & 0.442 (0.046) & 0.393 (0.041) & 0.0 (0.0) & 0.044 (0.014) \\
& 5 & 0.382 (0.074) & 0.477 (0.093) & 0.425 (0.082) & 0.058 (0.165) & 0.043 (0.033) \\
& 10 & 0.162 (0.052) & 0.203 (0.065) & 0.18 (0.058) & 0.031 (0.182) & 0.089 (0.022)\\ \hline
\end{tabular}%
}
\end{table}
\clearpage

\begin{figure}
\hspace*{-3cm}
\renewcommand{\figurename}{Figure S}
    \centering
    \includegraphics{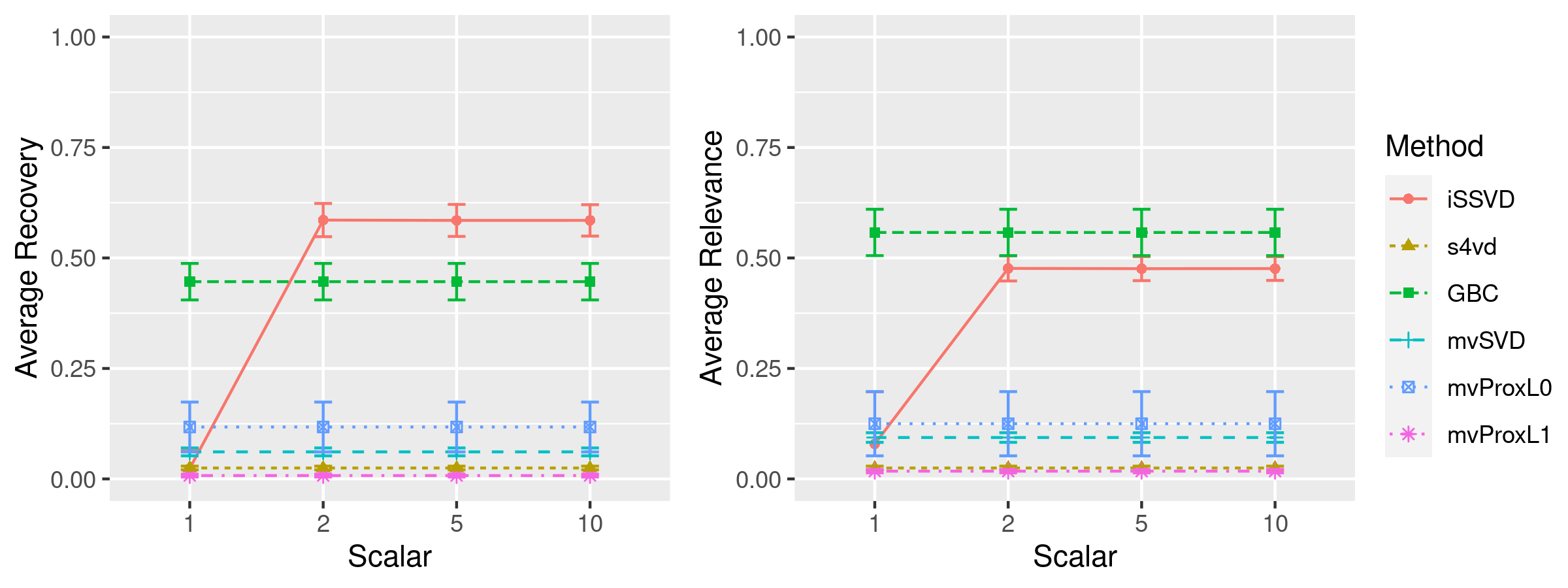}
    \caption{Scenario 1 case 1 when each variable is standardized to have mean zero and variance one. The performance of all algorithms is much lower
compared with un-standardized data. The relevance and recovery scores of iSSVD are greater than
others but only around 0.5. When scalar is 1 all scores of all algorithms are around 0.}
    \label{scalar_standard}
\end{figure}
\clearpage

\begin{figure}
\hspace*{-3cm}
\renewcommand{\figurename}{Figure S}
    \centering
    \includegraphics{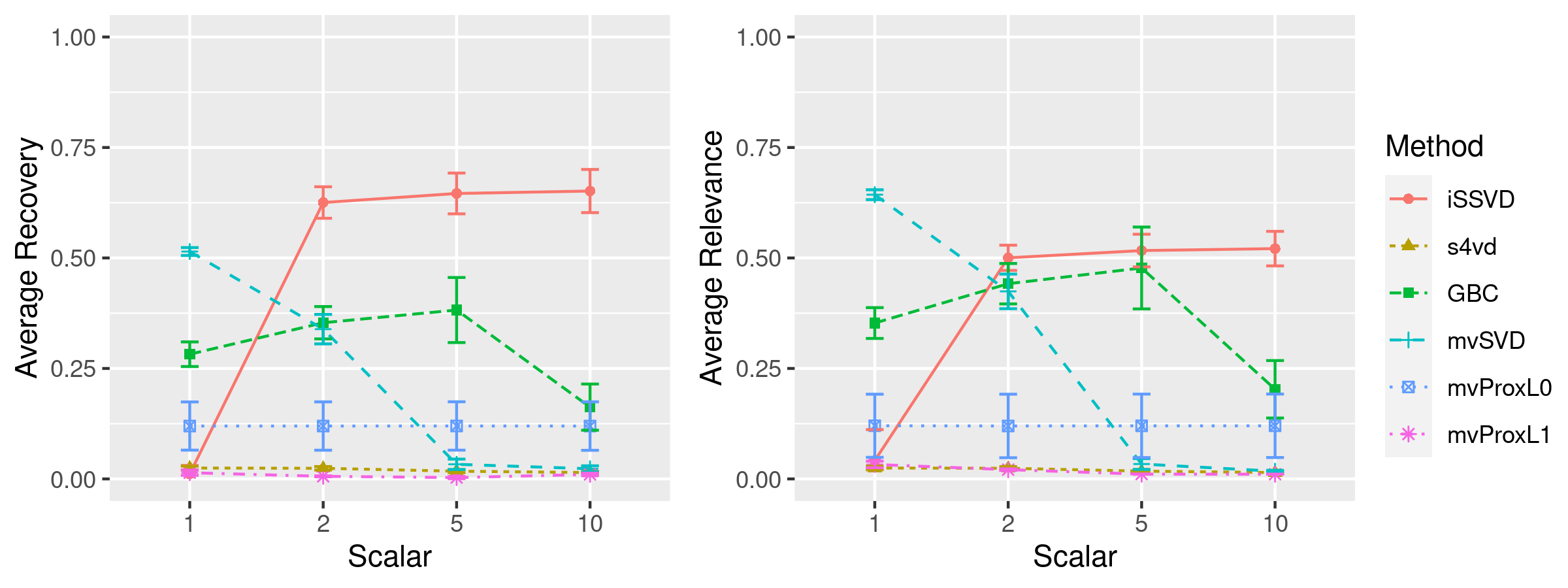}
    \caption{Scenario 1 case 1 when each variable is centered to have mean zero but not scaled to have variance one. The performance of all algorithms is
much lower compared with un-centered data. The relevance and recovery scores of iSSVD are greater
than others but only around 0.5. When scalar is 1 all scores of all algorithms but mvSVD are around 0.
}
    \label{fig:my_label}
\end{figure}
\clearpage

\begin{figure}
\hspace*{-3cm}
\renewcommand{\figurename}{Figure S}
    \centering
    \includegraphics{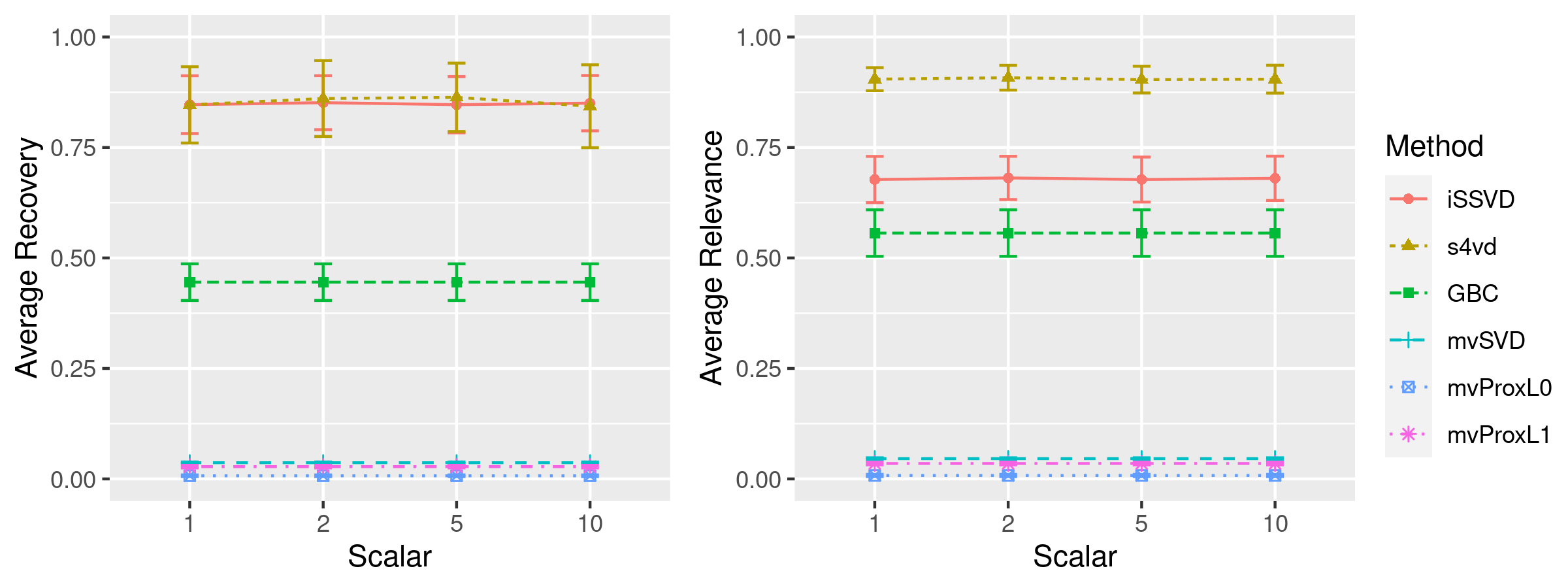}
    \caption{Scenario 1 case 1 when variables are each scaled to have variance one but not centered to have mean zero. The
performance of iSSVD is comparable to results from original data (Figure 2) in the main text.
text.
}
    \label{fig:my_label}
\end{figure}
\clearpage

\begin{figure}
\hspace*{-3cm}
\renewcommand{\figurename}{Figure S}
    \centering
    \includegraphics{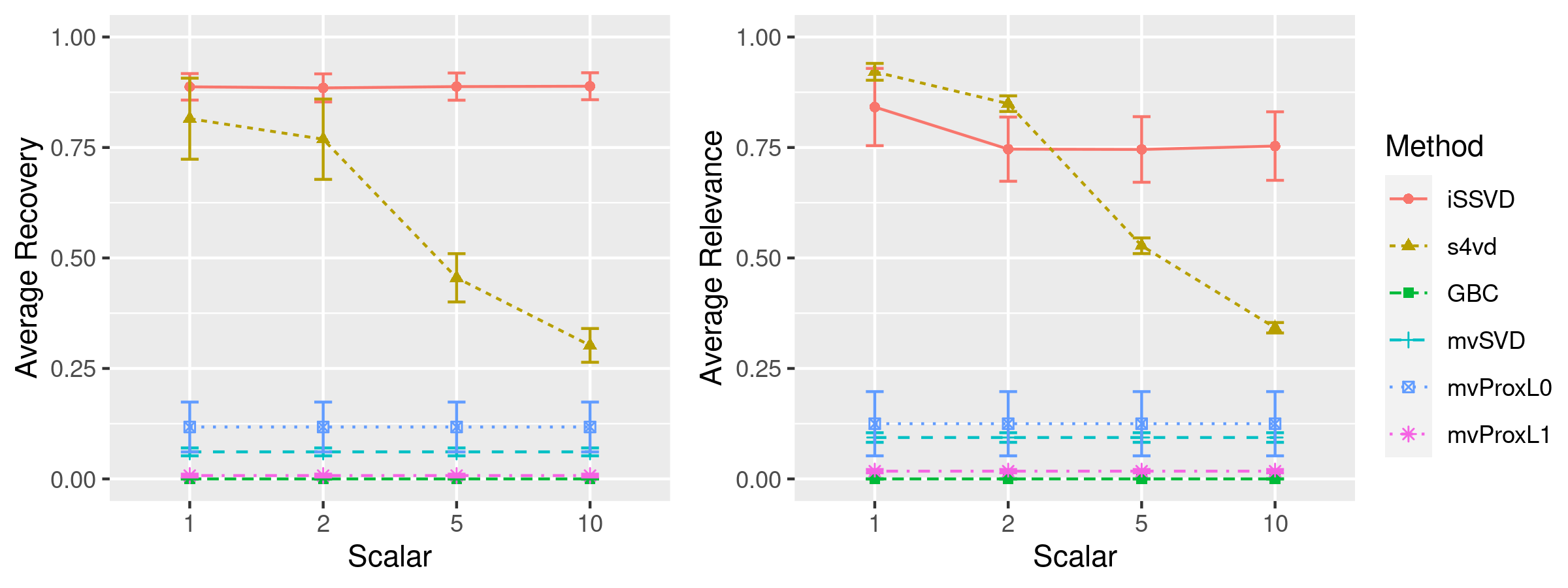}
    \caption{Scenario 1 case 1 when each View is scaled by their Frobenius norms, so that each View is normalized to have Frobenius norm 1. The
performance of iSSVD is comparable to results from original data (Figure 2) in the main text.
text.
}
    \label{fig:my_label}
\end{figure}
\clearpage

\begin{figure}
\hspace*{-3cm}
\renewcommand{\figurename}{Figure S}
    \centering
    \includegraphics{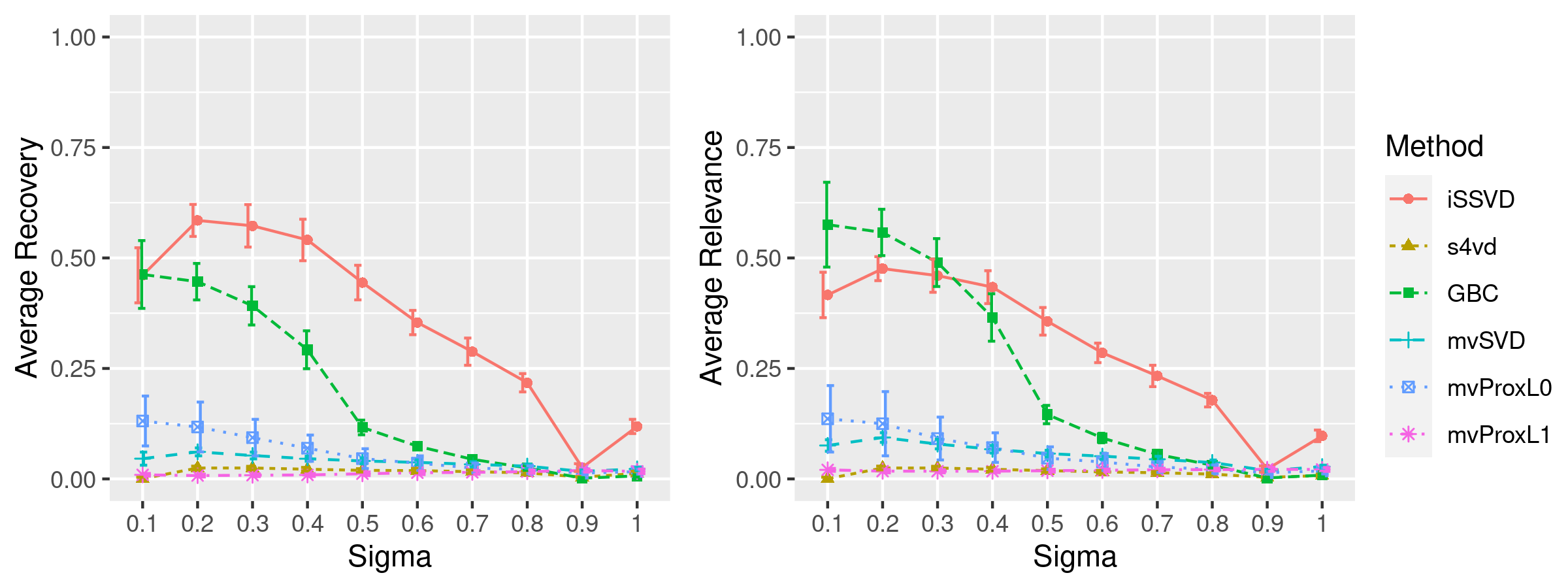}
    \caption{Scenario 1 case 2 when each variable is standardized to have mean zero and variance one. The performance of all algorithms is much lower compared with unstandardized data. }
    \label{sig_stand}
\end{figure}
\clearpage

\begin{figure}
\hspace*{-3cm}
\renewcommand{\figurename}{Figure S}
    \centering
    \includegraphics{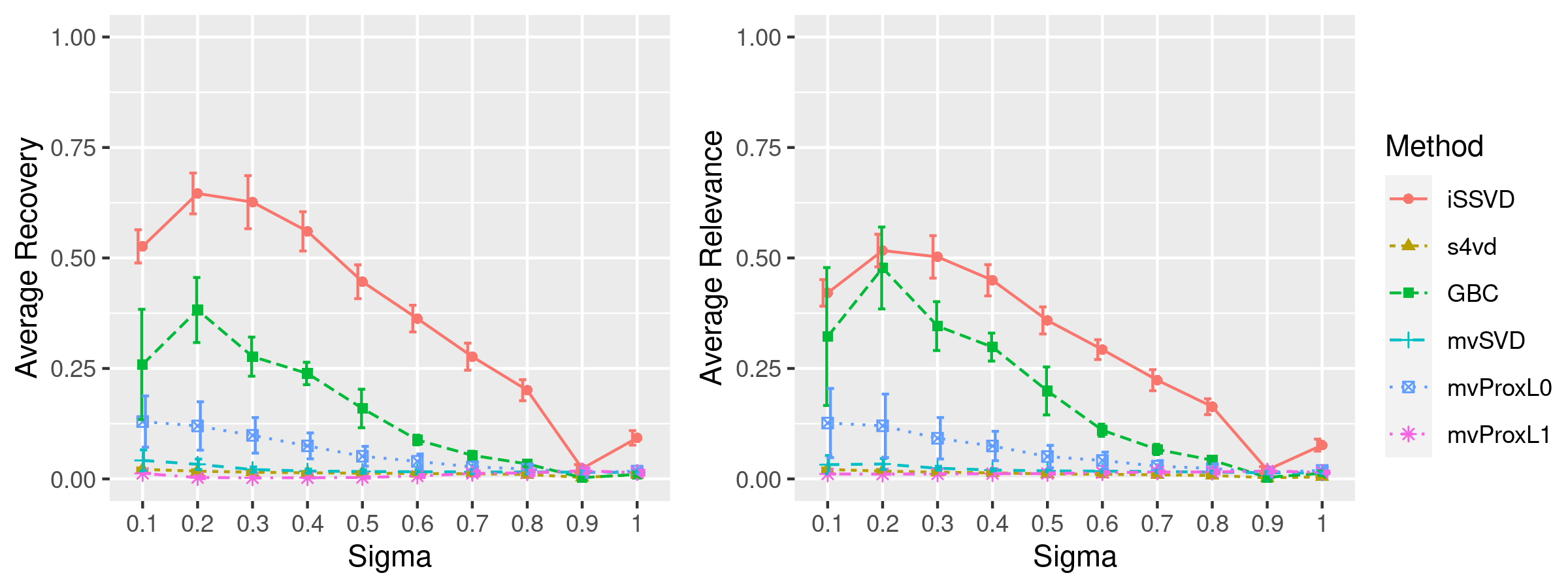}
    \caption{Scenario 1 case 2 when each variable is centered to have mean zero but not scaled to have variance 1. The performance of all algorithms is much lower compared with unstandardized data.}
    \label{sig_center}
\end{figure}
\clearpage

\begin{figure}
\hspace*{-3cm}
\renewcommand{\figurename}{Figure S}
    \centering
    \includegraphics{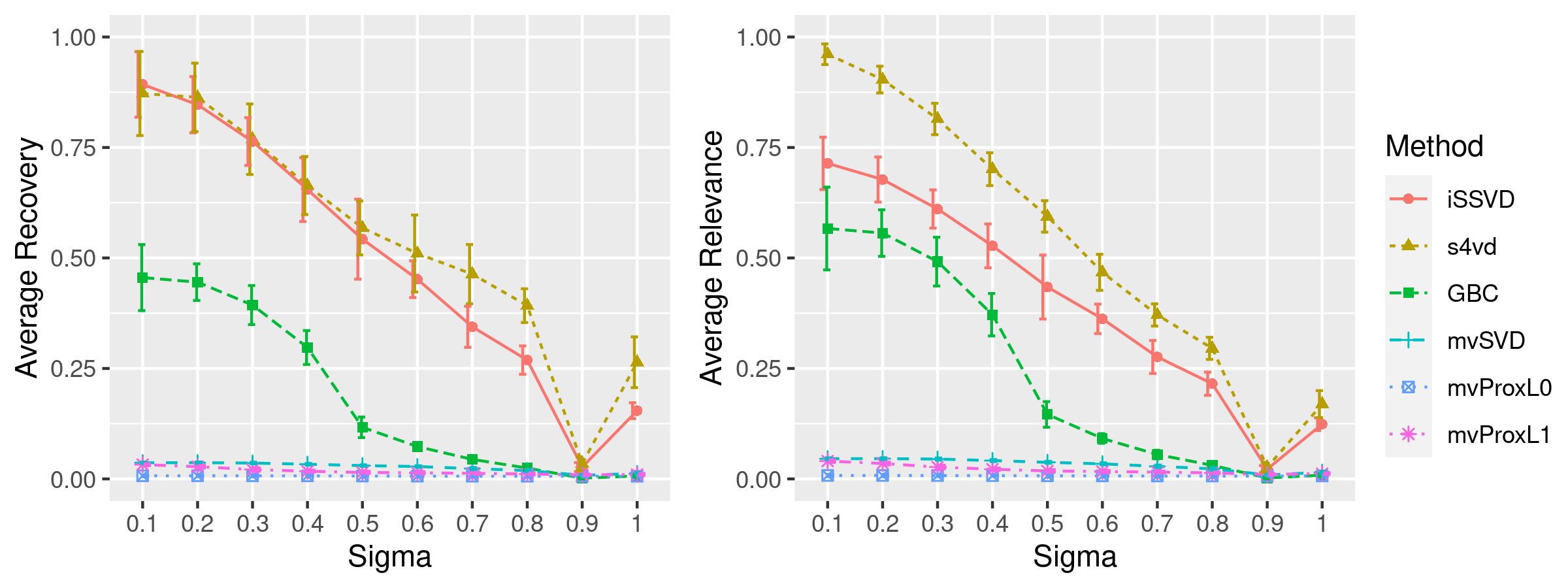}
    \caption{Scenario 1 case 2 when  variables are each scaled to have variance one but not centered to have mean zero. The
performance of iSSVD is comparable to results from original data (Figure 3) in the main text.
text. 
}
    \label{fig:my_label}
\end{figure}
\clearpage

\begin{figure}
\hspace*{-3cm}
\renewcommand{\figurename}{Figure S}
    \centering
    \includegraphics{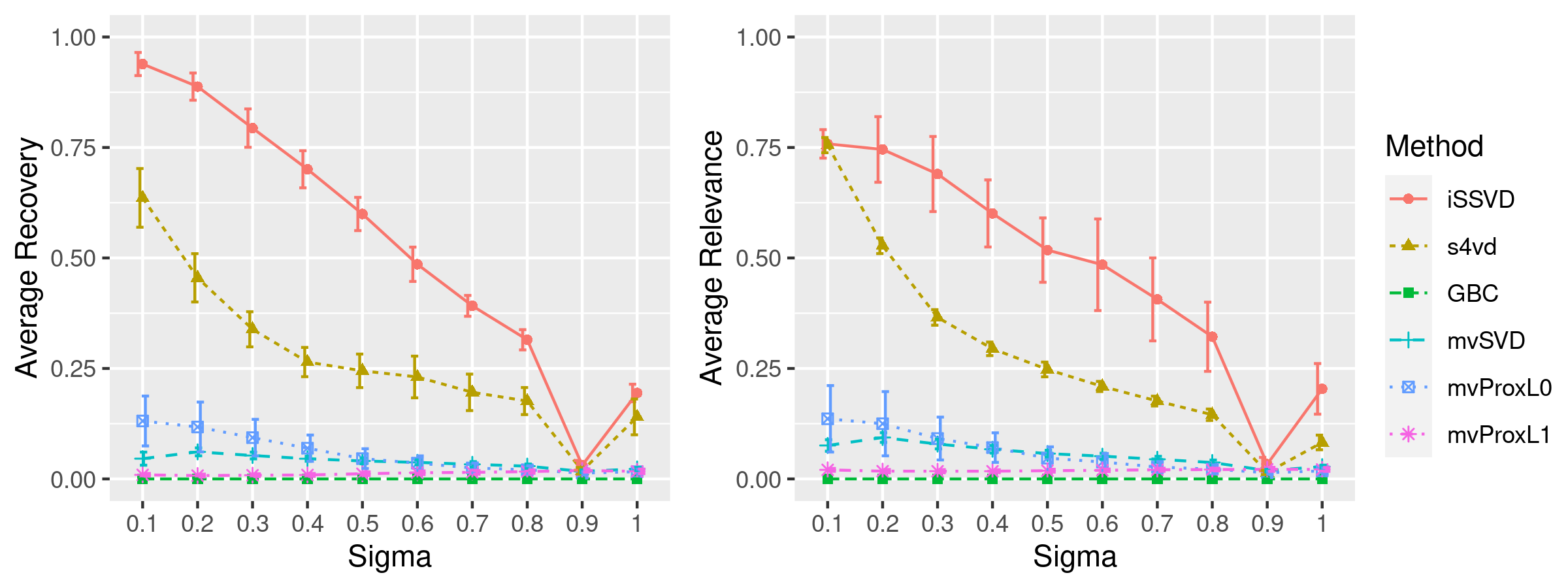}
    \caption{Scenario 1 case 2 when each View is scaled by their Frobenius norms, so that each View is normalized to have Frobenius norm 1. The
performance of iSSVD is comparable to results from original data (Figure 3) in the main text.
text.}
    \label{sig_center}
\end{figure}
\clearpage

\small
\begin{table}[]
\renewcommand\thetable{S1.5}
\centering
\caption{Parameter settings for simulations.}
\label{TableS14}
\resizebox{0.35\textwidth}{!}{%
\begin{tabular}{ll}
\hline
\textbf{Algorithms} & \textbf{Parameters} \\ \hline
\textbf{iSSVD} & standr = False \\
 & pointwise = True \\
 & steps = 100 \\
 & size = 0.5 \\
 & ssthr = {[}0.6,0.8{]} \\
 & nbicluster = 5 \\
 & rows\_nc = True \\
 & cols\_nc = True \\
 & col\_overlap = False \\
 & row\_overlap = False \\
 & pceru = 0.1 \\
 & pcerv = {[}0.1,0.1{]} \\
 & merr = 0.0001 \\
 & iters = 100 \\ \hline
\textbf{s4vd} & pointwise = True \\
 & steps = 100 \\
 & size = 0.5 \\
 & ssthr = c(0.6,0.8) \\
 & nbicluster = 5 \\
 & rows\_nc = True \\
 & cols\_nc = True \\
 & col\_overlap = False \\
 & row\_overlap = False \\
 & pceru = 0.1 \\
 & pcerv = 0.1 \\
 & merr = 0.0001 \\
 & iters = 100 \\ \hline
\textbf{mvSVD} & lvs = c(0.9, 0.9) \\
 & lz = 0.9 \\ \hline
\textbf{mvProxL0} & svs = c(110,110) \\
 & sz = 12 \\ \hline
\textbf{mvProxL1} & lus = c(0.7,0.7) \\
 & lvs = c(20,20) \\
 & lz = 0.8 \\ \hline
\textbf{GBC} & L=5 \\ & dist.type=rep(0,p) \\
& param=rep(0.25,p) \\ & use.network=FALSE \\ & maxiter=100 \\ & tol=0.001 \\ & nu1=5 \\ & n4=20 \\ & cutoff=20 \\ \hline
\end{tabular}%
}
\end{table}
\clearpage

\small
\begin{table}[]
\renewcommand\thetable{S1.6}
\centering
\caption{Parameter settings for real data analysis.}
\label{TableS14}
\resizebox{0.35\textwidth}{!}{%
\begin{tabular}{ll}
\hline
\textbf{Algorithms} & \textbf{Parameters} \\ \hline
\textbf{iSSVD} & standr = False \\
 & pointwise = True \\
 & steps = 100 \\
 & size = 0.6 \\
 & ssthr = {[}0.6,0.8{]} \\
 & nbicluster = 10 \\
 & rows\_nc = True \\
 & cols\_nc = True \\
 & col\_overlap = False \\
 & row\_overlap = False \\
 & pceru = 0.15 \\
 & pcerv = {[}0.16,0.01{]} \\
 & merr = 0.0001 \\
 & iters = 100 \\ \hline
\textbf{s4vd} & pointwise = T \\
 & steps = 100 \\
 & size = 0.6 \\
 & ssthr = c(0.6,0.8) \\
 & nbicluster = 10 \\
 & rows\_nc = T \\
 & cols\_nc = T \\
 & col\_overlap = F \\
 & row\_overlap = F \\
 & pceru = 0.2 \\
 & pcerv = 0.1 \\
 & merr = 0.0001 \\
 & iters = 100 \\ \hline
\textbf{mvSVD} & lvs = c(0.9, 0.9) \\
 & lz = 0.9 \\ \hline
\textbf{mvProxL0} & svs = c(110,130) \\
 & sz = 90 \\ \hline
\textbf{mvProxL1} & lus = c(0.5,0.5) \\
 & lvs = c(20,20) \\
 & lz = 0.5 \\ \hline
 \textbf{GBC} & L=5 \\ & dist.type=rep(0,3433) \\
& param=rep(0.25,3433) \\ & use.network=FALSE \\ & maxiter=100 \\ & tol=0.0001 \\ & nu1=5 \\ & n4=20 \\ & cutoff=20 \\ \hline
\end{tabular}%
}
\end{table}
\clearpage

\begin{table}[]
\renewcommand\thetable{S2.1}
\centering
\caption{Averaged unclustered samples. Values are mean (SD).Results are from Scenario 2.}
\label{TableS21}
\resizebox{\textwidth}{!}{%
\begin{tabular}{ccccccccccc}
\hline
\multicolumn{1}{l}{} & \multicolumn{10}{c}{\textbf{Sigma}} \\ \hline
 & 0.1 & 0.2 & 0.3 & 0.4 & 0.5 & 0.6 & 0.7 & 0.8 & 0.9 & 1 \\ \hline
iSSVD & 0 (0) & 8.34 (7.74) & 22.1 (2.79) & 45.2 (20.8) & 71.9 (22.4) & 85.6 (23.2) & 95.8 (19.5) & 98.3 (15.4) & 104 (17.9) & 115 (24.5) \\
s4vd & 151 (0) & 151 (0) & 145 (16.4) & 116 (29.2) & 100 (15.1) & 110 (16.6) & 107 (32.7) & 116 (40.3) & 135 (34.5) & 139 (39.6) \\
mvSVD & 52.6 (4.98) & 49.2 (5.23) & 41.3 (3.70) & 33.6 (4.87) & 27.8 (4.08) & 23.6 (3.03) & 19.7 (4.03) & 14.8 (3.35) & 8.6 (2.73) & 4.38 (2.33) \\
mvProxL0 & 140 (0) & 140 (0) & 140 (0) & 140 (0) & 140 (0) & 140 (0) & 140 (0) & 140 (0) & 140 (0) & 140 (0) \\
mvProxL1 & 100 (0) & 97 (12.0) & 51 (7.07) & 50.4 (0.835) & 50.4 (1.94) & 44.3 (5.36) & 29.0 (4.83) & 19.2 (3.69) & 10.8 (2.84) & 6.64 (2.41) \\ \hline
\end{tabular}%
}
\end{table}
\clearpage

\small
\renewcommand\thetable{S2.2}
\setlength\LTleft{-85pt}     
\begin{longtable}{@{\extracolsep{\fill}}lllllll@{}}
\caption{Simulation results of scenario 2. The scalar is 1, i.e. no unbalanced scales. Values are mean
(SD).}
\label{TableS22}\\
\hline
\textbf{} & \textbf{Sigma} & \textbf{Ave. Recovery} & \textbf{Ave. Relevace} & \textbf{F-scores} & \textbf{FP rate} & \textbf{FN rate} \\ \hline
\endfirsthead
\endhead
\hline
\endfoot
\endlastfoot
\textbf{iSSVD} & 0.1 & 0.927 (0.0118) & 0.927 (0.0118) & 0.927 (0.0118) & 0.0706 (0.0113) & 0.00958 (0.00346) \\
\textbf{} & 0.2 & 0.888 (0.0315) & 0.888 (0.0315) & 0.888 (0.0315) & 0.0688 (0.0108) & 0.0537 (0.0377) \\
\textbf{} & 0.3 & 0.805 (0.0133) & 0.805 (0.0133) & 0.805 (0.0133) & 0.0720 (0.00954) & 0.142 (0.0111) \\
\textbf{} & 0.4 & 0.653 (0.0820) & 0.677 (0.0539) & 0.662 (0.0656) & 0.0840 (0.0211) & 0.276 (0.0577) \\
\textbf{} & 0.5 & 0.490 (0.0742) & 0.536 (0.0705) & 0.509 (0.0641) & 0.103 (0.0314) & 0.426 (0.0753) \\
\textbf{} & 0.6 & 0.372 (0.0695) & 0.408 (0.0686) & 0.387 (0.0646) & 0.118 (0.0309) & 0.559 (0.0753) \\
\textbf{} & 0.7 & 0.286 (0.0532) & 0.321 (0.0655) & 0.300 (0.0536) & 0.131 (0.0352) & 0.649 (0.0721) \\
\textbf{} & 0.8 & 0.233 (0.0199) & 0.265 (0.0439) & 0.247 (0.0247) & 0.156 (0.0284) & 0.704 (0.0487) \\
\textbf{} & 0.9 & 0.0401 (0.00841) & 0.0466 (0.0110) & 0.0428 (0.00888) & 0.358 (0.0333) & 0.936 (0.0147) \\
\textbf{} & 1 & 0.122 (0.0201) & 0.169 (0.0486) & 0.124 (0.0170) & 0.195 (0.0389) & 0.805 (0.0541) \\ \hline
\textbf{s4vd} & 0.1 & 0.953 (0.0128) & 0.238 (0.00320) & 0.381 (0.00512) & 0.763 (0.0141) & 0.755 (0) \\
\textbf{} & 0.2 & 0.953 (0.0122) & 0.238 (0.00306) & 0.381 (0.00490) & 0.764 (0.0134) & 0.755 (0) \\
\textbf{} & 0.3 & 0.891 (0.143) & 0.258 (0.0613) & 0.387 (0.0259) & 0.684 (0.226) & 0.732 (0.0620) \\
\textbf{} & 0.4 & 0.649 (0.242) & 0.355 (0.0992) & 0.417 (0.0544) & 0.330 (0.347) & 0.635 (0.100) \\
\textbf{} & 0.5 & 0.400 (0.109) & 0.358 (0.0358) & 0.371 (0.0462) & 0.0639 (0.0969) & 0.631 (0.0376) \\
\textbf{} & 0.6 & 0.321 (0.0754) & 0.271 (0.0310) & 0.289 (0.0381) & 0.0771 (0.0650) & 0.719 (0.0328) \\
\textbf{} & 0.7 & 0.250 (0.109) & 0.196 (0.0553) & 0.226 (0.0279) & 0.0951 (0.102) & 0.735 (0.189) \\
\textbf{} & 0.8 & 0.213 (0.0907) & 0.139 (0.0456) & 0.177 (0.0199) & 0.120 (0.0817) & 0.773 (0.231) \\
\textbf{} & 0.9 & 0.0418 (0.0179) & 0.0219 (0.00755) & 0.0289 (0.00703) & 0.221 (0.104) & 0.932 (0.192) \\
\textbf{} & 1 & 0.166 (0.0696) & 0.0692 (0.0218) & 0.100 (0.0135) & 0.180 (0.0882) & 0.864 (0.221) \\ \hline
\textbf{mvSVD} & 0.1 & 0.565 (0.0222) & 0.690 (0.0207) & 0.622 (0.0201) & 0 (0) & 0.310 (0.0207) \\
\textbf{} & 0.2 & 0.489 (0.0208) & 0.587 (0.0175) & 0.534 (0.0179) & 0.0914 (0.0300) & 0.307 (0.0225) \\
\textbf{} & 0.3 & 0.313 (0.00978) & 0.362 (0.00938) & 0.335 (0.00914) & 0.416 (0.0619) & 0.291 (0.0174) \\
\textbf{} & 0.4 & 0.207 (0.00426) & 0.251 (0.00526) & 0.227 (0.00452) & 0.788 (0.110) & 0.271 (0.0225) \\
\textbf{} & 0.5 & 0.160 (0.00434) & 0.190 (0.00590) & 0.174 (0.00488) & 1.37 (0.179) & 0.289 (0.0229) \\
\textbf{} & 0.6 & 0.135 (0.00343) & 0.156 (0.00316) & 0.145 (0.00324) & 1.88 (0.207) & 0.310 (0.0219) \\
\textbf{} & 0.7 & 0.117 (0.00407) & 0.135 (0.00430) & 0.125 (0.00414) & 2.32 (0.216) & 0.338 (0.0252) \\
\textbf{} & 0.8 & 0.0928 (0.00734) & 0.106 (0.00825) & 0.0990 (0.00775) & 2.42 (0.371) & 0.415 (0.0412) \\
\textbf{} & 0.9 & 0.0304 (0.00246) & 0.0313 (0.00267) & 0.0308 (0.00253) & 2.20 (0.473) & 0.789 (0.0182) \\
\textbf{} & 1 & 0.0513 (0.00528) & 0.0569 (0.00641) & 0.0540 (0.00579) & 1.66 (0.372) & 0.584 (0.0381) \\ \hline
\textbf{mvProxL0} & 0.1 & 0.101 (0.0462) & 0.0597 (0.0170) & 0.0730 (0.0221) & 0.197 (0.0176) & 0.933 (0.0176) \\
\textbf{} & 0.2 & 0.0835 (0.0364) & 0.0499 (0.0154) & 0.0610 (0.0198) & 0.208 (0.0160) & 0.944 (0.0160) \\
\textbf{} & 0.3 & 0.0441 (0.0145) & 0.0342 (0.00964) & 0.0380 (0.0106) & 0.223 (0.0105) & 0.959 (0.0105) \\
\textbf{} & 0.4 & 0.0357 (0.0142) & 0.0285 (0.00888) & 0.0313 (0.0102) & 0.230 (0.00996) & 0.966 (0.00996) \\
\textbf{} & 0.5 & 0.0222 (0.00828) & 0.0197 (0.00530) & 0.0207 (0.00628) & 0.240 (0.00610) & 0.976 (0.00610) \\
\textbf{} & 0.6 & 0.0171 (0.00497) & 0.0158 (0.00374) & 0.0163 (0.00410) & 0.244 (0.00447) & 0.980 (0.00447) \\
\textbf{} & 0.7 & 0.0145 (0.00436) & 0.0137 (0.00377) & 0.0140 (0.00394) & 0.247 (0.00443) & 0.983 (0.00443) \\
\textbf{} & 0.8 & 0.0124 (0.00327) & 0.0123 (0.00250) & 0.0123 (0.00269) & 0.249 (0.00303) & 0.985 (0.00303) \\
\textbf{} & 0.9 & 0.0102 (0.00113) & 0.0101 (0.00118) & 0.0101 (0.00111) & 0.251 (0.00145) & 0.987 (0.00145) \\
\textbf{} & 1 & 0.0113 (0.00176) & 0.0109 (0.00172) & 0.0111 (0.00164) & 0.250 (0.00211) & 0.986 (0.00211) \\ \hline
\textbf{mvProxL1} & 0.1 & 0.220 (0.00617) & 0.275 (0.00772) & 0.245 (0.00686) & 0 (0) & 0.5 (0) \\
\textbf{} & 0.2 & 0.0809 (0.0118) & 0.101 (0.0148) & 0.0899 (0.0131) & 0 (0) & 0.485 (0.0600) \\
\textbf{} & 0.3 & 0.0859 (0.00428) & 0.107 (0.00534) & 0.0954 (0.00475) & 0 (0) & 0.255 (0.0354) \\
\textbf{} & 0.4 & 0.0747 (0.000498) & 0.0934 (0.000611) & 0.0830 (0.000545) & 0 (0) & 0.253 (0.00432) \\
\textbf{} & 0.5 & 0.0680 (0.000777) & 0.0855 (0.000832) & 0.0757 (0.000797) & 0 (0) & 0.264 (0.00982) \\
\textbf{} & 0.6 & 0.0656 (0.00769) & 0.0815 (0.00659) & 0.0727 (0.00718) & 0.0942 (0.277) & 0.296 (0.0221) \\
\textbf{} & 0.7 & 0.0638 (0.00451) & 0.0702 (0.00490) & 0.0668 (0.00460) & 1.20 (0.533) & 0.394 (0.0386) \\
\textbf{} & 0.8 & 0.0478 (0.00508) & 0.0513 (0.00527) & 0.0495 (0.00512) & 1.46 (0.423) & 0.465 (0.0315) \\
\textbf{} & 0.9 & 0.0287 (0.00220) & 0.0297 (0.00194) & 0.0292 (0.00203) & 1.20 (0.472) & 0.503 (0.0320) \\
\textbf{} & 1 & 0.0344 (0.00300) & 0.0369 (0.00320) & 0.0356 (0.00296) & 0.832 (0.379) & 0.429 (0.0319) \\ \hline
\textbf{GBC} & 0.1 & 0.141 (0.003) & 0.178 (0.003) & 0.158 (0.003) & 0.0 (0.0) & 0.007 (0.007) \\
\textbf{} & 0.2 & 0.124 (0.006) & 0.171 (0.008) & 0.144 (0.007) & 0.0 (0.0) & 0.21 (0.023) \\
\textbf{} & 0.3 & 0.097 (0.005) & 0.136 (0.004) & 0.113 (0.004) & 0.0 (0.0) & 0.588 (0.015) \\
\textbf{} & 0.4 & 0.082 (0.006) & 0.109 (0.007) & 0.093 (0.006) & 0.0 (0.0) & 0.788 (0.012) \\
\textbf{} & 0.5 & 0.058 (0.005) & 0.074 (0.006) & 0.065 (0.005) & 0.0 (0.0) & 0.899 (0.01) \\
\textbf{} & 0.6 & 0.033 (0.005) & 0.041 (0.006) & 0.037 (0.006) & 0.0 (0.0) & 0.954 (0.008) \\
\textbf{} & 0.7 & 0.014 (0.005) & 0.018 (0.006) & 0.016 (0.006) & 0.0 (0.0) & 0.981 (0.007) \\
\textbf{} & 0.8 & 0.004 (0.002) & 0.005 (0.002) & 0.004 (0.002) & 0.0 (0.0) & 0.995 (0.003) \\
\textbf{} & 0.9 & 0.001 (0.001) & 0.001 (0.001) & 0.001 (0.001) & 0.0 (0.0) & 0.999 (0.001) \\
\textbf{} & 1 & 0.001 (0.0) & 0.001 (0.001) & 0.001 (0.0) & 0.0 (0.0) & 0.999 (0.001)\\ \hline
\end{longtable}

\clearpage
\begin{table}[]
\renewcommand\thetable{S2.3}
\centering
\caption{Here we show the frequencies of the number of biclusters detected by iSSVD under different noise levels in scenario 2. Values are count (proportions).}
\label{detected}
\begin{tabular}{cccc}
\hline
 & \multicolumn{3}{c}{\textbf{Number of biclusters detected}} \\
\textbf{Sigma} & \textbf{less than 4} & \textbf{4} & \textbf{more than 4} \\ \hline
0.1 & 0 (0) & 50 (1) & 0 (0) \\
0.2 & 0 (0) & 27 (0.54) & 23 (0.46) \\
0.3 & 0 (0) & 23 (0.46) & 27 (0.54) \\
0.4 & 5 (0.1) & 16 (0.32) & 29 (0.58) \\
0.5 & 18 (0.36) & 11 (0.22) & 21 (0.42) \\
0.6 & 16 (0.32) & 9 (0.18) & 25 (0.50) \\
0.7 & 19 (0.38) & 12 (0.24) & 19 (0.38) \\
0.8 & 24 (0.48) & 9 (0.18) & 17 (0.34) \\
0.9 & 25 (0.50) & 6 (0.12) & 19 (0.38) \\
1 & 27 (0.54) & 8 (0.16) & 15 (0.30) \\ \hline
\end{tabular}
\end{table}
\clearpage

\subsection{Considering outliers in simulations}
We consider another set of simulations to assess the performance of the methods in the presence of potential outliers.  There are two views and  four biclusters. Each bicluster has $50$ samples and $250$ variables, thus there are $n=200$ samples and $p^{(1)}=1,000$ variables for View 1. Similar for View 2. The data for View 1 was simulated as $\mathbf{X}^{(1)} = \widetilde{\mathbf{X}}^{(1)} + \mathbf{E}^{(1)}$. Here, $\widetilde{\mathbf{X}}^{(1)}$ is of size $200 \times 1,000$ and is block-diagonal with four blocks (representing biclusters) each of size $50 \times 250$. The entries for the first 5 rows (samples) for bicluster 1 are random integers from 0 and 1, inclusive and the entries for the remaining 45 rows are random integers from 1 to 2, inclusive. The entries for bicluster 2 are generated as random integers from 1 to 3 inclusive; entries from bicluster 3 are generated as random integers from 2 and 3 inclusive. For bicluster 4, entries for 48 rows are generated as random integers from 0 to 2 inclusive, and the entries for samples 199 and 200 in bicluster 4 are set as 1.1 and 1.2 respectively. The off-diagonal entries in $\widetilde{\mathbf{X}}^{(1)}$ are all set to zeros. The error term $\mathbf{E}^{(1)}$ is a matrix of size $200 \times 1,000$ and the entries are generated from the normal distribution with mean $0$ and standard deviation $0.1$. Data for View 2, $\mathbf{X}^{(2)}$,  are generated similarly, but the entries for the error term are from the normal distribution with mean $0$ and standard deviation $0.15$.

\begin{figure}[htb!]
\begin{tabular}{cc}
         \centering
         \includegraphics[width=.5\textwidth]{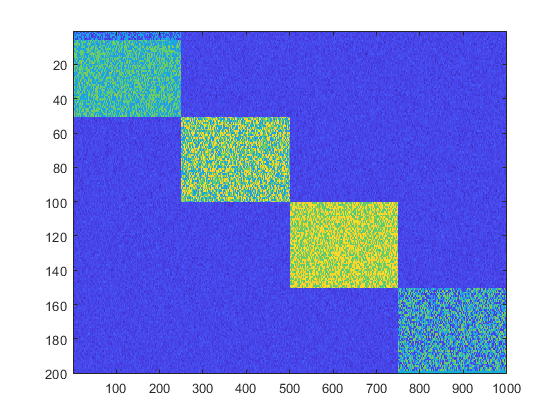}& 
         \includegraphics[width=.5\textwidth]{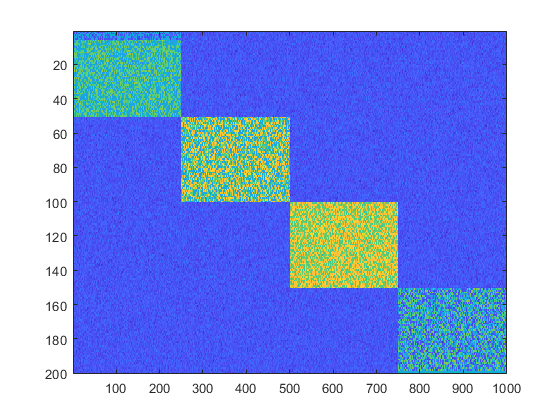}\\
          \includegraphics[width=.5\textwidth]{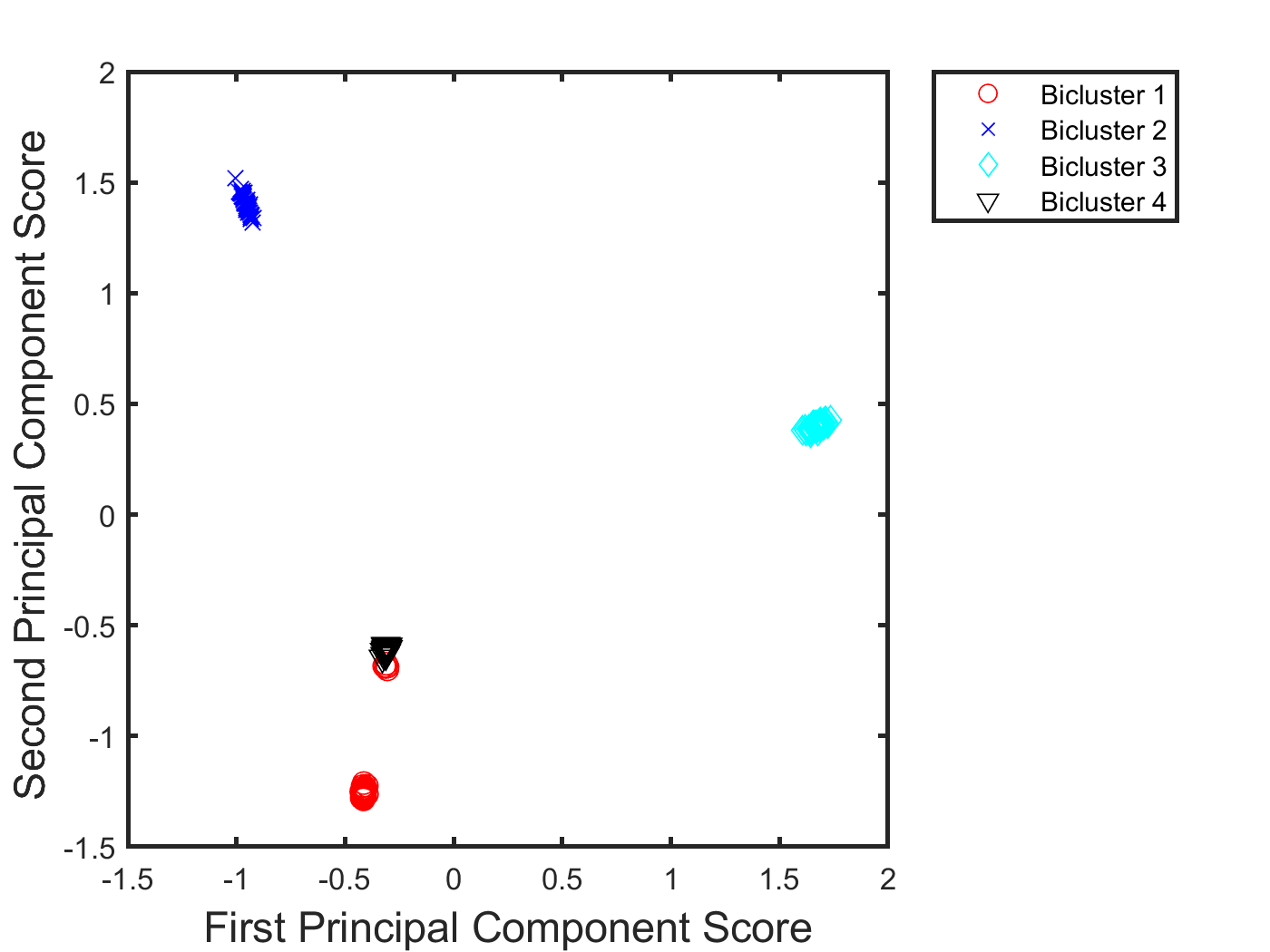}&\includegraphics[width=.5\textwidth]{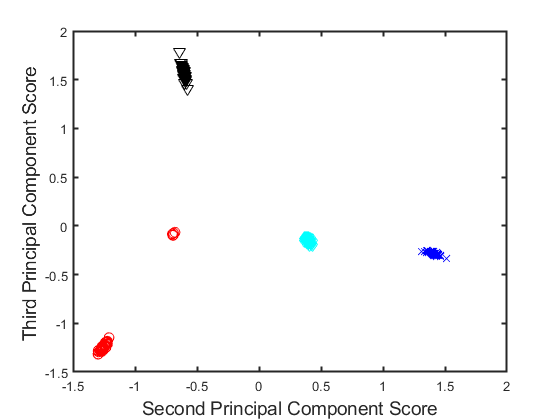}\\
          \includegraphics[width=.5\textwidth]{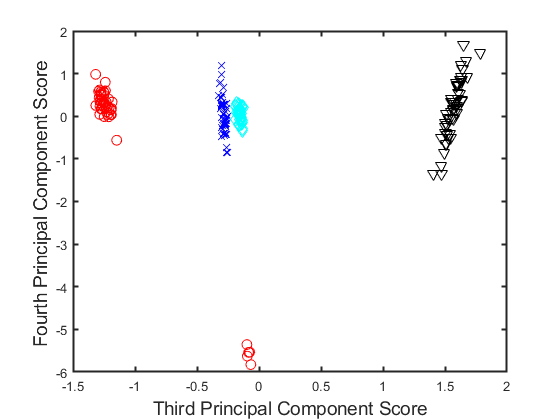}&\includegraphics[width=.5\textwidth]{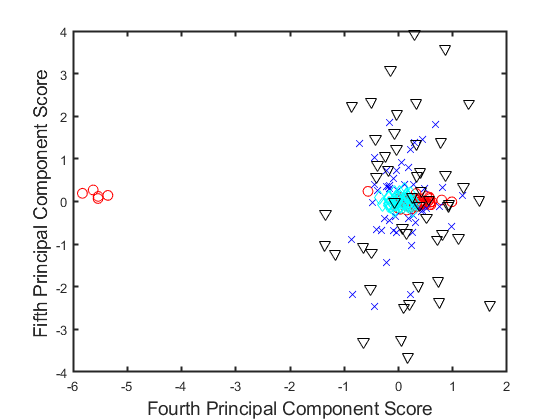}\\
\end{tabular}
    \caption{Upper Panels: Image plot of simulated data for View 1 (right) and View 2 (left); y-axis for samples and x-axis for variables. Middle and Lower Panels: Plots of principal component scores from principal component analysis. Some samples for Bicluster 1 [colored red and denoted as circles] are far from the center of samples for Bicluster 1. This is apparent from all four principal component plots. These samples may be considered as outliers.}
    \label{fig:outliers}
\end{figure}

\begin{figure}[ht!]
    \centering
    \includegraphics[scale=0.7]{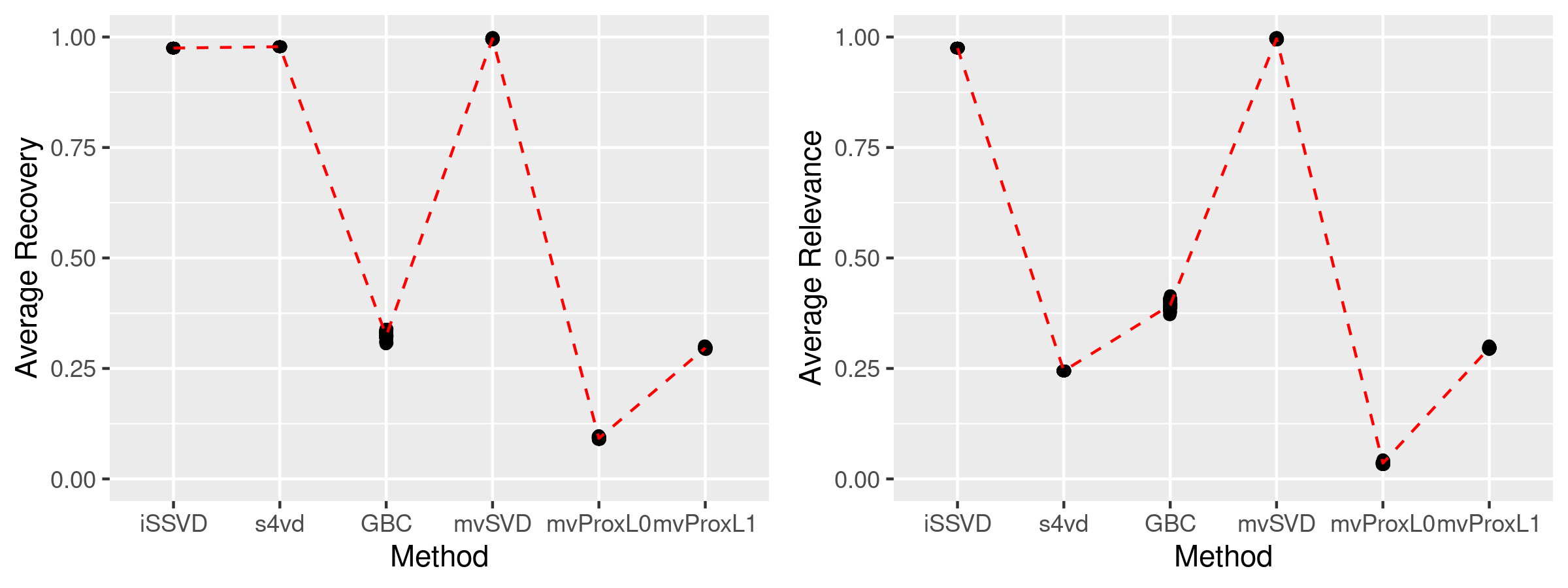}
    \caption{Simulation results for the outliers scenario. }
    \label{fig:outlier}
\end{figure}

Figure \ref{fig:outliers} shows image plots for Views 1 and 2 and principal component (PC) score plots from principal component analysis of View 1 data. The PC scores plot show that some samples from bicluster 1 (colored red and denoted as circles) are farther away from the center of the samples in that bicluster, and may be taken as potential outliers. Based on Figure \ref{fig:outlier} and Table S3.1, the proposed method, iSSVD has average recovery and relevance estimates that are comparable to mvSVD and better than s4vd, GBC, mvProxL0 and mvProxL1. 

\begin{table}[ht!]
\renewcommand\thetable{S3.1}
\caption{Simulation results of outliers case.}
\label{tab:outlierresults}
\resizebox{\textwidth}{!}{%
\begin{tabular}{llllll}
\hline
\textbf{Method} & \multicolumn{1}{c}{\textbf{Ave. Recovery}} & \multicolumn{1}{c}{\textbf{Ave. Relevance}} & \multicolumn{1}{c}{\textbf{F-scores}} & \multicolumn{1}{c}{\textbf{FP rate}} & \multicolumn{1}{c}{\textbf{FN rate}} \\
\hline
\textbf{iSSVD} & 0.975 (0.0) & 0.975 (0.0) & 0.975 (0.0) & 0.0 (0.0) & 0.025 (0.0) \\
\textbf{s4vd} & 0.978 (0.0) & 0.245 (0.0) & 0.391 (0.0) & 0.734 (0.0) & 0.755 (0.0) \\
\textbf{mvSVD} & 0.996 (0.001) & 0.996 (0.001) & 0.996 (0.001) & 0.004 (0.001) & 0.0 (0.0) \\
\textbf{mvProxL0} & 0.092 (0.002) & 0.036 (0.003) & 0.051 (0.003) & 0.069 (0.003) & 0.964 (0.003) \\
\textbf{mvProxL1} & 0.296 (0.002) & 0.296 (0.002) & 0.296 (0.002) & 2.385 (0.017) & 0.0 (0.0) \\
\textbf{GBC} & 0.324 (0.009) & 0.393 (0.009) & 0.355 (0.009) & 0.034 (0.014) & 0.0 (0.0)\\
\hline
\hline
\end{tabular}%
}
\end{table}
\section{More real data analysis} 
In this section we show the supplementary results of the COPD data analysis. Compared to iSSVD, other competing methods (except mvProxL0) did not find clinically meaningful subgroups from the data. The sample clusters detected by the other methods did not show any differences in demographic, clinical, and key COPD outcome variables (Tables S4.2 - S4.4). The sample clusters detected by \textit{mvProxL0} showed differences in some variables, but were not different across key lung function variables such as FEV$_1$/FVC and FEV$_1$ [\% predicted]). Compared to \textit{mvProxL0}, the sample clusters detected by our method showed differences in more variables. Five-year follow-up data were available for $162$ individuals. We observe that the subgroups identified by iSSVD are again well-differentiated on some key outcomes, clinical variables and symptoms (Table S4.5). In addition to clinical characteristics, we show pathways and disease  enrichment analysis  in Tables S4.6 and S4.7.

\begin{table}[]
\renewcommand\thetable{S4.1}
\resizebox{\textwidth}{!}{%
\begin{tabular}{lllllll}
\hline
\hline
\multicolumn{1}{c}{\textbf{Bicluster   (N)}} & \multicolumn{1}{c}{\textbf{1 (92)}} & \multicolumn{1}{c}{\textbf{2 (92)}} & \multicolumn{1}{c}{\textbf{3 (93)}} & \multicolumn{1}{c}{\textbf{4 (93)}} & \multicolumn{1}{c}{\textbf{5 (93)}} & \multicolumn{1}{c}{\textbf{P-value}} \\ \hline
\hline
\textbf{Demographics and Clinical} & \multicolumn{1}{c}{\textbf{}} & \multicolumn{1}{c}{\textbf{}} & \multicolumn{1}{c}{\textbf{}} & \multicolumn{1}{c}{\textbf{}} & \multicolumn{1}{c}{\textbf{}} & \multicolumn{1}{c}{\textbf{}} \\
Age (years) & 67.65 (8.61) & 67.25 (8.28) & 68.23 (8.07) & 68.37 (7.88) & 65.4 (9.3) & 0.1135 \\
Body Mass Index & 28.89 (5.97) & 28.9 (5.88) & 29.32 (6.57) & 29.03 (6.18) & 29.64 (6.75) & 0.9285 \\
Pack Years & 39.06 (18.86) & 42.06 (27.23) & 45.86 (25.71) & 48.28 (23.59) & 50.3 (26.96) & 0.0062 \\
Duration of smoking (years) & 34.73 (12.7) & 35.86 (11.65) & 37.39 (12.31) & 36.99 (11.42) & 39.19 (11.59) & 0.1603 \\
Gender &  &  &  &  &  &  \\
~~~~Males & 46 (50\%) & 39 (42\%) & 44 (47\%) & 50 (54\%) & 56 (60\%) & 0.1524 \\
~~~~Females & 46 (50\%) & 53 (58\%) & 49 (53\%) & 43 (46\%) & 37 (40\%) &  \\
Smoking Status &  &  &  &  &  &  \\
~~~~Former & 66 (72\%) & 78 (85\%) & 73 (78\%) & 76 (82\%) & 59 (63\%) & 0.005 \\
~~~~Current & 26 (28\%) & 14 (15\%) & 20 (22\%) & 17 (18\%) & 34 (37\%) &  \\
Dyspnea Score (MMRC) & 0.57 (1.04) & 1.15 (1.25) & 1.29 (1.45) & 1.27 (1.36) & 1.16 (1.29) & 0.0006 \\
BODE Index & 0.78 (1.26) & 1.33 (1.98) & 1.75 (2.48) & 1.99 (2.33) & 1.67 (2.02) & 0.0054 \\
\textbf{Outcomes} &  &  &  &  &  &  \\
FEV1/FVC & 0.7 (0.13) & 0.67 (0.15) & 0.64 (0.17) & 0.63 (0.18) & 0.67 (0.14) & 0.1037 \\
FEV1 (\% predicted) & 84.01 (21.23) & 77.93 (24.77) & 76.81 (27.35) & 72.83 (26.5) & 77.93 (25.8) & 0.1708 \\
COPD Status &  &  &  &  &  &  \\
~~~~No & 48 (58\%) & 37 (46\%) & 37 (43\%) & 37 (44\%) & 37 (46\%) & 0.3144 \\
~~~~Yes & 35 (42\%) & 43 (54\%) & 49 (57\%) & 47 (56\%) & 43 (54\%) &  \\
\textbf{Symptoms} &  &  &  &  &  &  \\
Exacerbation Frequency & 0.14 (0.48) & 0.27 (0.74) & 0.29 (0.68) & 0.12 (0.39) & 0.23 (0.78) & 0.169 \\
Percent Emphysema (Thirona) & 4.0 (6.5) & 8.12 (10.29) & 8.92 (13.43) & 9.47 (11.8) & 5.74 (8.95) & 0.0232 \\
Percent 15 & -920.03 (19.39) & -928.38 (25.26) & -927.16 (28.3) & -929.61 (27.5) & -923.1 (24.12) & 0.0773 \\
Ever had Asthma &  &  &  &  &  &  \\
~~~~No & 90 (98\%) & 91 (99\%) & 90 (97\%) & 90 (97\%) & 89 (96\%) & 0.7334 \\
~~~~Yes & 2 (2\%) & 1 (1\%) & 3 (3\%) & 3 (3\%) & 4 (4\%) &  \\
Gastroesophageal Reflux &  &  &  &  &  &  \\
~~~~No & 66 (72\%) & 54 (59\%) & 64 (69\%) & 59 (63\%) & 65 (70\%) & 0.3144 \\
~~~~Yes & 26 (28\%) & 38 (41\%) & 29 (31\%) & 34 (37\%) & 28 (30\%) &  \\
\multicolumn{2}{l}{Been to ER   or hospitalized for lung problems} &  &  &  &  &  \\
~~~~No & 89 (97\%) & 87 (95\%) & 79 (85\%) & 89 (96\%) & 84 (90\%) & 0.0139 \\
~~~~Yes & 3 (3\%) & 5 (5\%) & 14 (15\%) & 4 (4\%) & 9 (10\%) &  \\ 
\hline
\hline
\end{tabular}%
}
\caption{Clinical characteristics of mvProxL0 biclusters of COPD data. Values are mean for continuous variables, and N  for binary/categorical variables. P-values are from comparing all clusters using Kruskal-Wallis test for continuous variables, and Chi-square test for categorical variables. MMRC- modified Medical Research Council dyspnea score (0-4, 4= most severe symptoms). GOLD stages 1-4 are combined to form COPD cases; GOLD stage 0 form COPD controls.}
\label{tab:my-table}
\end{table}

\begin{table}[]
\renewcommand\thetable{S4.2}
\centering
\begin{scriptsize}
\begin{tabular}{llll}
\hline
\hline
\multicolumn{1}{c}{\textbf{Bicluster   (N)}} & \multicolumn{1}{c}{\textbf{1 (455)}} & \multicolumn{1}{c}{\textbf{2 (8)}} & \multicolumn{1}{c}{\textbf{P-value}} \\ \hline
\hline
\textbf{Demographics and Clinical} & \multicolumn{1}{c}{\textbf{}} & \multicolumn{1}{c}{\textbf{}} & \multicolumn{1}{c}{\textbf{}} \\
Age (years) & 67.4 (8.51) & 66.04 (6.33) & 0.6457 \\
Body Mass Index & 29.17 (6.3) & 28.33 (3.35) & 0.9405 \\
Pack Years & 45.22 (24.91) & 40.31 (25.07) & 0.5163 \\
Duration of smoking (years) & 36.93 (11.98) & 31.68 (11.95) & 0.272 \\
Gender &  &  &  \\
~~~~Males & 231 (51\%) & 4 (50\%) & 0.7539 \\
~~~~Females & 224 (49\%) & 4 (50\%) &  \\
Smoking Status &  &  &  \\
~~~~Former & 345 (76\%) & 7 (88\%) & 0.727 \\
~~~~Current & 110 (24\%) & 1 (12\%) &  \\
Dyspnea Score (MMRC) & 1.1 (1.31) & 0.62 (1.19) & 0.2425 \\
BODE Index & 1.52 (2.1) & 0.88 (1.46) & 0.4017 \\
\textbf{Outcomes} &  &  &  \\
FEV1/FVC & 0.66 (0.15) & 0.7 (0.11) & 0.6823 \\
FEV1 (\% predicted) & 77.77 (25.36) & 84.62 (27.0) & 0.3529 \\
COPD Status &  &  &  \\
~~~~No & 191 (47\%) & 5 (71\%) & 0.3685 \\
~~~~Yes & 215 (53\%) & 2 (29\%) &  \\
\textbf{Symptoms} &  &  &  \\
Exacerbation Frequency & 0.21 (0.64) & 0.0 (0.0) & 0.2636 \\
Percent Emphysema (Thirona) & 7.32 (10.7) & 3.8 (5.1) & 0.6238 \\
Percent 15 & -925.71 (25.4) & -923.12 (17.3) & 0.8719 \\
Ever had Asthma &  &  &  \\
~~~~No & 443 (97\%) & 7 (88\%) & 0.5522 \\
~~~~Yes & 12 (3\%) & 1 (12\%) &  \\
Gastroesophageal Reflux &  &  &  \\
~~~~No & 304 (67\%) & 4 (50\%) & 0.5345 \\
~~~~Yes & 151 (33\%) & 4 (50\%) &  \\
\multicolumn{2}{l}{Been to ER   or hospitalized for lung problems} &  &  \\
~~~~No & 420 (92\%) & 8 (100\%) & 0.8876 \\
~~~~Yes & 35 (8\%) & 0 (0\%) &  \\ 
\hline
\hline
\end{tabular}%
\end{scriptsize}
\caption{Clinical characteristics of mvProxL1 biclusters. Values are mean for continuous variables, and N  for binary/categorical variables. P-values are from comparing all clusters using Kruskal-Wallis test for continuous variables, and Chi-square test for categorical variables. MMRC- modified Medical Research Council dyspnea score (0-4, 4= most severe symptoms). GOLD stages 1-4 are combined to form COPD cases; GOLD stage 0 form COPD controls.}
\label{tab:my-table}
\end{table}

\begin{table}[]
\renewcommand\thetable{S4.3}
\resizebox{\textwidth}{!}{%
\begin{tabular}{llllll}
\hline
\hline
\multicolumn{1}{c}{\textbf{Bicluster   (N)}} & \multicolumn{1}{c}{\textbf{1 (390)}} & \multicolumn{1}{c}{\textbf{2 (64)}} & \multicolumn{1}{c}{\textbf{3 (6)}} & \multicolumn{1}{c}{\textbf{4 (3)}} & \multicolumn{1}{c}{\textbf{P-value}} \\ \hline
\hline
\textbf{Demographics and Clinical} & \multicolumn{1}{c}{\textbf{}} & \multicolumn{1}{c}{\textbf{}} & \multicolumn{1}{c}{\textbf{}} & \multicolumn{1}{c}{\textbf{}} & \multicolumn{1}{c}{\textbf{}} \\
Age (years) & 67.44 (8.46) & 67.2 (8.67) & 66.22 (9.36) & 65.33 (7.75) & 0.9721 \\
Body Mass Index & 28.97 (6.12) & 30.21 (7.11) & 28.03 (3.25) & 33.32 (8.31) & 0.3289 \\
Pack Years & 45.23 (23.87) & 45.6 (31.24) & 37.8 (20.77) & 37.2 (17.64) & 0.7043 \\
Duration of smoking (years) & 37.12 (11.79) & 35.98 (13.25) & 30.47 (8.65) & 31.53 (15.36) & 0.4072 \\
Gender &  &  &  &  &  \\
~~~~Males & 195 (50\%) & 35 (55\%) & 5 (83\%) & 0 (0\%) & 0.1057 \\
~~~~Females & 195 (50\%) & 29 (45\%) & 1 (17\%) & 3 (100\%) &  \\
Smoking Status &  &  &  &  &  \\
~~~~Former & 289 (74\%) & 55 (86\%) & 6 (100\%) & 2 (67\%) & 0.0989 \\
~~~~Current & 101 (26\%) & 9 (14\%) & 0 (0\%) & 1 (33\%) &  \\
Dyspnea Score (MMRC) & 1.11 (1.31) & 1.05 (1.34) & 0.67 (0.82) & 0.0 (0.0) & 0.3801 \\
BODE Index & 1.52 (2.06) & 1.61 (2.38) & 0.33 (0.52) & 0.33 (0.58) & 0.4072 \\
\textbf{Outcomes} &  &  &  &  &  \\
FEV1/FVC & 0.66 (0.15) & 0.66 (0.16) & 0.75 (0.07) & 0.75 (0.08) & 0.4733 \\
FEV1 (\% predicted) & 77.65 (25.71) & 77.37 (24.04) & 87.02 (15.82) & 102.27 (14.73) & 0.2839 \\
COPD Status &  &  &  &  &  \\
~~~~No & 166 (47\%) & 25 (45\%) & 3 (75\%) & 2 (67\%) & 0.6252 \\
~~~~Yes & 185 (53\%) & 30 (55\%) & 1 (25\%) & 1 (33\%) &  \\
\textbf{Symptoms} &  &  &  &  &  \\
Exacerbation Frequency & 0.23 (0.67) & 0.12 (0.42) & 0.0 (0.0) & 0.0 (0.0) & 0.4522 \\
Percent Emphysema (Thirona) & 7.14 (10.43) & 8.56 (12.28) & 3.24 (4.84) & 2.67 (2.82) & 0.4507 \\
Percent 15 & -925.52 (24.79) & -927.61 (28.88) & -919.67 (17.75) & -915.33 (22.85) & 0.6717 \\
Ever had Asthma &  &  &  &  &  \\
~~~~No & 378 (97\%) & 63 (98\%) & 6 (100\%) & 3 (100\%) & 0.8668 \\
~~~~Yes & 12 (3\%) & 1 (2\%) & 0 (0\%) & 0 (0\%) &  \\
Gastroesophageal Reflux &  &  &  &  &  \\
~~~~No & 261 (67\%) & 42 (66\%) & 3 (50\%) & 2 (67\%) & 0.8526 \\
~~~~Yes & 129 (33\%) & 22 (34\%) & 3 (50\%) & 1 (33\%) &  \\
Been to ER or hospitalized for lung   problems &  &  &  &  &  \\
~~~~No & 360 (92\%) & 59 (92\%) & 6 (100\%) & 3 (100\%) & 0.861 \\
~~~~Yes & 30 (8\%) & 5 (8\%) & 0 (0\%) & 0 (0\%) &  \\ 
\hline
\hline
\end{tabular}%
}
\caption{Clinical characteristics of mvSVD biclusters of COPD data.  Values are mean (SD)for continuous variables, and N (percentages) for binary/categorical variables.  P-values arefrom comparing all four groups using Kruskal-Wallis test for continuous variables, and Chi-square test for categorical variables.  MMRC- modified Medical Research Council dyspneascore  (0-4,  4=  most  severe  symptoms).   GOLD  stages  1-4  are  combined  to  form  COPDcases; GOLD stage 0 form COPD controls.}
\label{tab:my-table}
\end{table}

\begin{table}[]
\renewcommand\thetable{S4.4}
\resizebox{\textwidth}{!}{%
\begin{tabular}{lccccc}
\hline
\hline
\multicolumn{1}{c}{\textbf{Bicluster   (N)}} & \multicolumn{1}{c}{\textbf{1 (143)}} & \multicolumn{1}{c}{\textbf{2 (109)}} & \multicolumn{1}{c}{\textbf{3 (115)}} & \multicolumn{1}{c}{\textbf{4 (96)}} & \multicolumn{1}{c}{\textbf{P-value}} \\ \hline
\hline
\textbf{Demographics and Clinical} & \multicolumn{1}{c}{\textbf{}} & \multicolumn{1}{c}{\textbf{}} & \multicolumn{1}{c}{\textbf{}} & \multicolumn{1}{c}{\textbf{}} & \multicolumn{1}{c}{\textbf{}} \\
Age (years) & 68.22 (8.18) & 67.66 (8.14) & 67.33 (8.43) & 65.87 (9.22) & 0.217 \\
Body Mass Index & 28.96 (5.29) & 29.06 (6.22) & 29.1 (6.23) & 29.64 (7.6) & 0.9978 \\
Pack Years & 45.53 (25.46) & 46.9 (26.39) & 42.59 (22.48) & 45.57 (25.23) & 0.7574 \\
Duration of smoking (years) & 36.07 (12.16) & 37.41 (12.63) & 36.84 (12.12) & 37.33 (10.89) & 0.8972 \\
Gender &  &  &  &  &  \\
~~~~Males & 78 (55\%) & 63 (58\%) & 51 (44\%) & 43 (45\%) & 0.1005 \\
~~~~Females & 65 (45\%) & 46 (42\%) & 64 (56\%) & 53 (55\%) &  \\
Smoking Status &  &  &  &  &  \\
~~~~Former & 113 (79\%) & 85 (78\%) & 86 (75\%) & 68 (71\%) & 0.4844 \\
~~~~Current & 30 (21\%) & 24 (22\%) & 29 (25\%) & 28 (29\%) &  \\
Dyspnea Score (MMRC) & 1.13 (1.28) & 1.12 (1.35) & 1.02 (1.32) & 1.07 (1.3) & 0.8344 \\
BODE Index & 1.34 (1.81) & 1.74 (2.38) & 1.48 (2.08) & 1.54 (2.16) & 0.9433 \\
\textbf{Outcomes} &  &  &  &  &  \\
FEV1/FVC & 0.66 (0.15) & 0.65 (0.16) & 0.67 (0.15) & 0.68 (0.16) & 0.5043 \\
FEV1 (\% predicted) & 79.11 (24.29) & 77.25 (26.62) & 76.32 (24.8) & 78.67 (26.44) & 0.6755 \\
COPD Status &  &  &  &  &  \\
~~~~No & 61 (47\%) & 45 (45\%) & 48 (48\%) & 42 (51\%) & 0.881 \\
~~~~Yes & 70 (53\%) & 54 (55\%) & 53 (52\%) & 40 (49\%) &  \\
\textbf{Symptoms} &  &  &  &  &  \\
Exacerbation Frequency & 0.23 (0.78) & 0.15 (0.43) & 0.17 (0.48) & 0.3 (0.74) & 0.379 \\
Percent Emphysema (Thirona) & 7.49 (9.68) & 8.96 (12.97) & 5.68 (9.01) & 6.86 (10.7) & 0.1283 \\
Percent 15 & -928.22 (23.27) & -928.43 (27.87) & -921.5 (24.96) & -923.69 (24.93) & 0.0422 \\
Ever had Asthma &  &  &  &  &  \\
~~~~No & 139 (97\%) & 107 (98\%) & 110 (96\%) & 94 (98\%) & 0.668 \\
~~~~Yes & 4 (3\%) & 2 (2\%) & 5 (4\%) & 2 (2\%) &  \\
Gastroesophageal Reflux &  &  &  &  &  \\
~~~~No & 91 (64\%) & 74 (68\%) & 74 (64\%) & 69 (72\%) & 0.5508 \\
~~~~Yes & 52 (36\%) & 35 (32\%) & 41 (36\%) & 27 (28\%) &  \\
Been to ER or hospitalized for lung   problems &  &  &  &  &  \\
~~~~No & 134 (94\%) & 99 (91\%) & 108 (94\%) & 87 (91\%) & 0.6721 \\
~~~~Yes & 9 (6\%) & 10 (9\%) & 7 (6\%) & 9 (9\%) &  \\ 
\hline
\hline
\end{tabular}%
}
\caption{Clinical characteristics of GBC biclusters of COPD data.  Values are mean (SD)for continuous variables, and N (percentages) for binary/categorical variables.  P-values arefrom comparing all four groups using Kruskal-Wallis test for continuous variables, and Chi-square test for categorical variables.  MMRC- modified Medical Research Council dyspneascore  (0-4,  4=  most  severe  symptoms).   GOLD  stages  1-4  are  combined  to  form  COPDcases; GOLD stage 0 form COPD controls.}
\label{tab:my-table}
\end{table}

\clearpage
\begin{table}[]
\renewcommand\thetable{S4.5}
\centering
\resizebox{\textwidth}{!}{%
\begin{tabular}{lccccc}
\hline
\hline
Bicluster (N) & 1 (36) & 2 (39) & 3(53) & 4 (34) & P-value \\
\hline
\hline
\textbf{Demographics and Clinical} & \\
Age (years) & 75.62 (5.63)	&68.74 (8.71)	&69.45 (8.22)&	71.84 (8.66) &  0.0014 \\
Body Mass Index & 28.17 (4.80)	&29.45 (6.78)&	29.10 (6.54)	&27.76 (6.76) &  0.654 \\
Pack Years & 54.34 (26.11)	&37.35 (18.50)&	42.78 (25.36)	&45.84 (28.62) &  0.0506 \\
Gender & &\\
~~~~Males & 21 (58\%)&	19 (51\%)&	23 (45\%)	&14 (41\%)&	 0.4796\\
~~~~Females &15 (42\%)&	18 (49\%)&	28 (55\%)&	20 (59\%)&	 \\
Smoking Status & &\\
~~~Never &  	1 (2.8\%)&	0 (0\%)	&0 (0\%)&	0 (0\%)&0.1419\\
~~~~Former & 32 (89\%)	&27 (73\%)&	38 (75\%)&	27 (79\%)&	 \\
~~~~Current&  3 (8.3\%)	&10 (27\%)	&13 (25\%)	&7 (21\%)&	  \\
Dyspnea Score (MMRC)  & 1.47 (1.50) & 1.14 (1.38) & 1.18 (1.37) & 1.15 (1.31) &  0.7215 \\
BODE Index & 2.42 (2.58) & 1.30 (2.13) & 1.61 (2.32) & 1.56 (1.76) &  0.168\\
 \textbf{Outcomes} & &\\
FEV$_{1}$/FVC & 0.60 (0.16)&	0.70(0.11)	&0.67(0.14)	&0.62(0.14) &  0.03299 \\
FEV$_1$ (\% predicted) & 73.06(29.54)	&83.54 (21.92)&	81.20 (27.13)	&75.95 (25.04) &  0.352 \\
COPD status& & &\\
~~~~No	&10 (29\%)&	19 (51\%)&	23 (49\%)&	11 (35\%)&	0.1722\\
~~~~Yes	&24 (71\%)&	18 (49\%)&	24 (51\%)&	20 (65\%)&	\\
\textbf{Symptoms} & & \\
Exacerbation Frequency & 0.33 (0.99) & 0.16 (0.99) & 0.29 (1.01) & 0.24 (0.55) & 0.862 \\
Percent Emphysema (Thirona) & 14.29 (10.88)	&6.33 (6.63)	&10.10 (11.93)	&10.67 (7.34) & 0.0005\\
Percent 15 & -942.42 (21.92)&	-923.86 (23.60)&	932.29 (23.83)&	-937.94 (18.51) & 0.0045\\
Asthma since last visit & &\\
~~~~No & 35 (97.2\%)&	37 (100\%)&	51 (100\%)&	34 (97.1\%)&	 0.3408 \\
~~~~Yes & 1 (2.8\%)&	0 (0\%)&	0 (0\%)&	1 (2.9\%)&	 \\
 Gastroesophageal Reflux & &\\
~~~~No & 18 (50\%) & 21 (57\%) & 24 (47\%) & 12 (35\%) &  0.4383 \\
 ~~~~Yes & 18 (50\%)&	16 (43\%)&	27 (53\%)&	22 (65\%)&	 \\
Been to ER or hospitalised for lung problems & &\\
~~~~No & 31 (86\%) & 36 (97.3\%) & 44 (86\%) & 28 (82\%) &  0.1454 \\
~~~~Yes & 5 (14\%)&	1 (2.7\%)&	7 (14\%)&	6 (18\%) &   \\
\hline
\hline
\end{tabular}%
}
\caption{Clinical characteristics of iSSVD biclusters of COPD follow-up data. Values are mean (SD) for continuous variables, and N (proportions) for binary/categorical variables. P-values are from comparing all four groups using Kruskal-Wallis test for continuous variables, and Chi-square test for categorical variables. MMRC- modified Medical Research Council dyspnea score (0-4, 4= most severe symptoms). GOLD stages 1-4 are combined to form COPD cases; GOLD stage 0 form COPD controls.}\label{tab:Table1}
\end{table}

\begin{landscape}
\begin{table}[]
\renewcommand\thetable{S4.6}
\centering
\begin{scriptsize}
\begin{tabular}{lll|ll}
\hline
\hline
 & Ingenuity Canonical Pathways & P-values & Diseases and Disorders & P-value range  \\
  \hline
 \hline
Bicluster 1 &  EIF2 Signaling & 2.23E-05 &  Cancer & 8.74E-03-4.61E-10 \\
 & IL-8 Signaling & 8.18E-05 &  Organismal Injury and Abnormalities  & 8.91E-03-4.61E-10 \\
 & mTOR Signaling & 8.48E-05 &  Gastrointestinal Disease  & 8.46E-03 - 3.99E-08 \\
 & Thrombopoietin  Signaling & 1.94E-04 & Inflammatory Response& 8.46E-03 - 1.47E-07\\
  & ErbB4 Signaling & 2.79E-04&  Tumor Morphology&8.46E-03 - 2.45E-06  \\
  \hline
Bicluster 2 & Ethanol Degradation IV & 6.07E-04 &  Cancer  & 1.77E-02-5.60E-09  \\
 & Oxidative Ethanol Degradation III & 6.97E-04 &  Organismal Injury and Abnormalities & 1.77E-02-5.60E-09\\
 & Phagosome Mutation & 6.28E-03 & Gastrointestinal Disease  & 1.53E-02 - 7.39E-06 \\
 & Taurine Biosynthesis & 1.45E-02 &  Reproductive System Disease & 1.77E-02 -5.22E-05\\
 & mTOR Signaling & 1.98E-02 & Developmental Disorder & 1.45E-02 - 5.29E-05 \\
 \hline
Bicluster 3 & Oxidative Phosphorylation & 5.47E-04 &Cancer & 2.94E-02 - 2.13E-07 \\
 & Estrogen Receptor Signaling &8.44E-04&  Organismal Injury and Abnormalities & 2.94E-02 - 2.13E-07 \\
 & Mitochondrial Dysfunction & 3.89E-03 & Endocrine System Disorders& 2.82E-02 - 1.90E-04 \\
 & Mismatch Repair in Eukaryotes & 4.17E-03 &  Metabolic Disease & 2.82E-02 - 1.90E-04 \\
 & Cell Cycle Control of Chromosomal Replication & 4.83E-03 &  Gastrointestinal Disease& 2.82E-02 - 2.10E-04 \\
 \hline
Bicluster 4 & Mitochondrial Dysfunction & 5.96E-04 &  Cancer & 2.17E-02 - 4.16E-09\\
 & Oxidative Phosphorylation & 4.35E-03 &  Organismal Injury and Abnormalities & 2.40E-02 - 4.16E-09 \\
 & Cell Cycle Control of Chromosomal Replication & 4.74E-03 & Hereditary Disorder &1.80E-02 - 1.39E-05 \\
 & Inflammasome pathway & 6.41E-03&  Neurological Disease& 2.22E-02 - 1.39E-05 \\
 & Pyridoxal 5'-phosphate Salvage Pathway & 7.49E-03 &  Metabolic Disease &2.40 E-02 - 1.29 E-04 \\
 \hline
 \hline
\end{tabular}
\caption{Top 5 Enriched IPA canonical pathways and diseases  of the genes characterizing each bicluster.}
\end{scriptsize}
\label{Table2}
\end{table}
\end{landscape}

\begin{landscape}
\begin{table}[]
\renewcommand\thetable{S4.7}
\centering
\begin{scriptsize}
\begin{tabular}{lll|lll}
\hline
 & Ingenuity Canonical Pathways & P-values& Diseases and Disorders & P-value range  \\
  \hline
Bicluster 1 & Coagulation System & 4.60E-10&  Cardiovascular Disease& 2.44E-04 - 1.92E-11  \\ 
 & Glioma Invasiveness Signaling & 8.31E-05 & Organismal Injury and Abnormalities & 2.64E-04 - 1.92E-11\\
 &Phagosome Mutation & 8.22E-04 & Inflammatory Disease & 7.90E-05 - 1.81E-10\\
 & Intrinsic Prothrombin Activation Pathway & 1.09E-03 &  Connective Tissue Disorders & 2.09E-04 - 7.55E-09  \\
 & Melatonin Degradation III & 1.17E-03  & Skeletal and Muscular Disorders & 2.09E-04 - 7.55E-09 \\
 \hline
Bicluster 2 & Cardiac Hypertrophy Signaling (Enhanced) & 6.13E-08& Dermatological Diseases and Conditions & 4.46E-04 - 3.43E-11 \\
 & Airway Pathology in Chronic Obstructive Pulmonary Disease & 1.54E-07  & Organismal Injury and Abnormalities & 6.24E-04-3.43E-11  \\
 & Role of Macrophages, Fibroblasts and Endothelial Cells in Rhematoid Arthritis & 1.03E-06  & Connective Tissue Disorders & 6.14E-04 - 2.17E-09\\
 & IL-17 Signaling & 1.51E-06 &Cancer & 4.89E-04 - 6.72E-09 \\
 & Regulation of the Epithelial Mesenchymal Transition by Growth Factors Pathway & 1.72E-06 &Immunological Disease & 6.06E-04 - 6.72E-09 \\
 \hline
Bicluster 3 & Coagulation System & 4.54E-08& Cardiovascular Disease & 7.62E-04-1.73E-09 \\
 & Airway Pathology in Chronic Obstructive Pulmonary Disease & 6.28E-06  & Organismal Injury and Abnormalities & 8.23E-04 - 1.73E-09  \\
 & Clathrin-mediated Endocytosis Signaling & 4.35E-05  & Inflammatory Response & 6.70E-04 - 4.70E-08\\
 & Role of Osteoblasts, Osteoclasts and Chondrocytes in Rhematoid Arthritis & 7.77E-05 &Immunological Disease & 8.23E-04 - 6.29E-08 \\
 & Systemic Lupus Erythematosus in B Cell Signaling Pathway & 1.76E-04 &Inflammtory Disease &4.41E-04 - 6.49E-08 \\
 \hline
Bicluster 4 & Role of Pattern Recognition Receptors in Recognition of Bacteria and Viruses & 2.27E-04 & Organismal Injury and Abnormalities & 5.43E-03-8.28E-06  \\
 & IL-17 Signaling & 3.86E-04 &  Inflammatory Response & 5.43E-03 - 1.53E-05 \\
 &  Bladder Cancer Signaling & 3.61E-03 & Neurological Disease  & 4.31E-03 - 1.59E-05\\ 
 & Airway Pathology in Chronic Obstructive Pulmonary Disease & 3.73E-03 & Hematological Disease   & 5.43E-03 - 1.91E-05 \\
 & LXR/RXR Activation & 6.02E-03  & Infectious Disease  & 5.43E-03 - 1.91E-05 \\
\hline
\hline
\end{tabular}
\caption{Top 5 Significantly Enriched IPA canonical pathways and diseases of the proteins characterizing each bicluster.}
\label{tab:my-table}
\end{scriptsize}
\end{table}
\end{landscape}